\documentclass{aa}

\usepackage{graphicx}
\usepackage{txfonts}
\usepackage{color}
\usepackage{xspace}
\usepackage{siunitx}
\usepackage{amsmath}
\usepackage{amssymb}
\usepackage{xcolor}
\usepackage{dashbox}
\usepackage{framed}
\usepackage{ocg-p}

\usepackage{hyperref}

\hypersetup{
    colorlinks=true,						    
	breaklinks=true,
	linkcolor=blue,							    
    citecolor=blue,							    
	filecolor=blue,							    
	urlcolor=blue,
	unicode=false,							    
	pdftoolbar=true,						    
	pdfmenubar=true,						    
	pdffitwindow=false,						    
	pdfstartview={Fit},						    
	pdftitle={VISION - Vienna Survey in Orion},	
	pdfauthor={Stefan Meingast},			    
	pdfsubject={Infrared extinction in Orion A},
	pdfcreator={Stefan Meingast},			    
	pdfkeywords={},
	pdfnewwindow=true,						    
	pdfdisplaydoctitle=true					    
}

\makeatletter
\renewcommand*\aa@pageof{, page \thepage{} of \pageref*{LastPage}}
\makeatother

\definecolor{stefan}{rgb}{0.86, 0.08, 0.24}

\definecolor{joao}{rgb}{0.01, 0.75, 0.24}

\definecolor{marco}{rgb}{0.0, 0.5, 1.0}

\newcommand{\pnicer}{\textsc{Pnicer}\xspace}
\newcommand{\nicer}{\textsc{Nicer}\xspace}
\newcommand{\nicest}{\textsc{Nicest}\xspace}
\newcommand{\oriona}{Orion~A\xspace}
\newcommand{\orionb}{Orion~B\xspace}

\newcommand{\cdbox}[1]{%
  \colorlet{currentcolor}{.}%
  {\color{blue}%
    \dbox{\color{currentcolor}#1}}%
}
\newcommand{\ToggleLayer}[2]{%
  \leavevmode
  \pdfstartlink user {
    /Subtype /Link
    /Border [0 0 0]%
    /A <<
      /S/JavaScript
      /JS (
         var aOCGs = this.getOCGs(), Layer;
         var Layers = "#1".split(","), Active = -1, i, l;
         for (l=0; l<Layers.length; l++) {
           Layer = Layers[l];
           for (i=0; aOCGs && i<aOCGs.length; i++) {
             if (aOCGs[i].state && aOCGs[i].name == Layer) {
               Active = l;
               aOCGs[i].state = false;
             }
           }
           if (Active >= 0) break;
         }
         if (Active == -1) {
           for (l=0; l<Layers.length; l++) {
             if (Layers[l] == "") Active = l;
           }
         }
         Active = Active + 1;
         if (Active == Layers.length) Active = 0;
         Layer = Layers[Active];
         for (i=0; aOCGs && i<aOCGs.length; i++) {
           if (aOCGs[i].name == Layer) aOCGs[i].state = true;
         }
      )
    >>
  }#2%
  \pdfendlink
}

\begin{document}

\defcitealias{meingast16}{Paper~I}

\title{VISION - Vienna Survey in Orion}
\subtitle{II. Infrared extinction in Orion A\thanks{Extinction maps are available at the CDS via anonymous ftp to \href{http://cdsarc.u-strasbg.fr}{cdsarc.u-strasbg.fr} (\href{ftp://130.79.128.5}{130.79.128.5}) or via \href{http://cdsweb.u-strasbg.fr/cgi-bin/qcat?J/A+A/}{cdsweb.u-strasbg.fr/cgi-bin/qcat?J/A+A/}}}

\author{Stefan Meingast\inst{1}
		\and Jo\~ao Alves\inst{1}
        \and Marco Lombardi\inst{2}
        }

\institute{Department of Astrophysics, University of Vienna, T\"urkenschanzstrasse 17, 1180 Wien, Austria
    	   \and University of Milan, Department of Physics, via Celoria 16, 20133 Milan, Italy
        }

\date{Received 19 June 2017 / Accepted ...}

\abstract{
We have investigated the shape of the extinction curve in the infrared up to \SI{{\sim}25}{\micro\metre} for the \oriona star-forming complex. The basis of this work is near-infrared data acquired with the Visual and Infrared Survey Telescope for Astronomy, in combination with Pan-STARRS and mid-infrared \textit{Spitzer} photometry. We obtain colour excess ratios for eight passbands by fitting a series of colour-colour diagrams. The fits are performed using Markov chain Monte Carlo methods, together with a linear model under a Bayesian formalism. The resulting colour excess ratios are directly interpreted as a measure of the extinction law. We show that the \oriona molecular cloud is characterized by flat mid-infrared extinction, similar to many other recently studied sightlines. Moreover, we find statistically significant evidence that the extinction law from \SI{{\sim}1}{\micro\metre} to at least \SI{{\sim}6}{\micro\metre} varies across the cloud. In particular, we find a gradient along galactic longitude, where regions near the Orion Nebula Cluster show a different extinction law compared to L1641 and L1647, the low-mass star-forming sites in the cloud complex. These variations are of the order of only $3\%$ and are most likely caused by the influence of the massive stars on their surrounding medium. While the observed general trends in our measurements are in agreement with model predictions, both well-established and new dust grain models are not able to fully reproduce our infrared extinction curve. We also present a new extinction map featuring a resolution of \SI{1}{\arcmin} and revisit the correlation between extinction and dust optical depth. This analysis shows that cloud substructure, which is not sampled by background sources, affects the conversion factor between these two measures. In conclusion, we argue that specific characteristics of the infrared extinction law are still not well understood, but \oriona can serve as an unbiased template for future studies.}


\maketitle

\section{Introduction}
\label{sec:introduction}

\begin{figure*}
    \centering
    \includegraphics[width=\hsize]{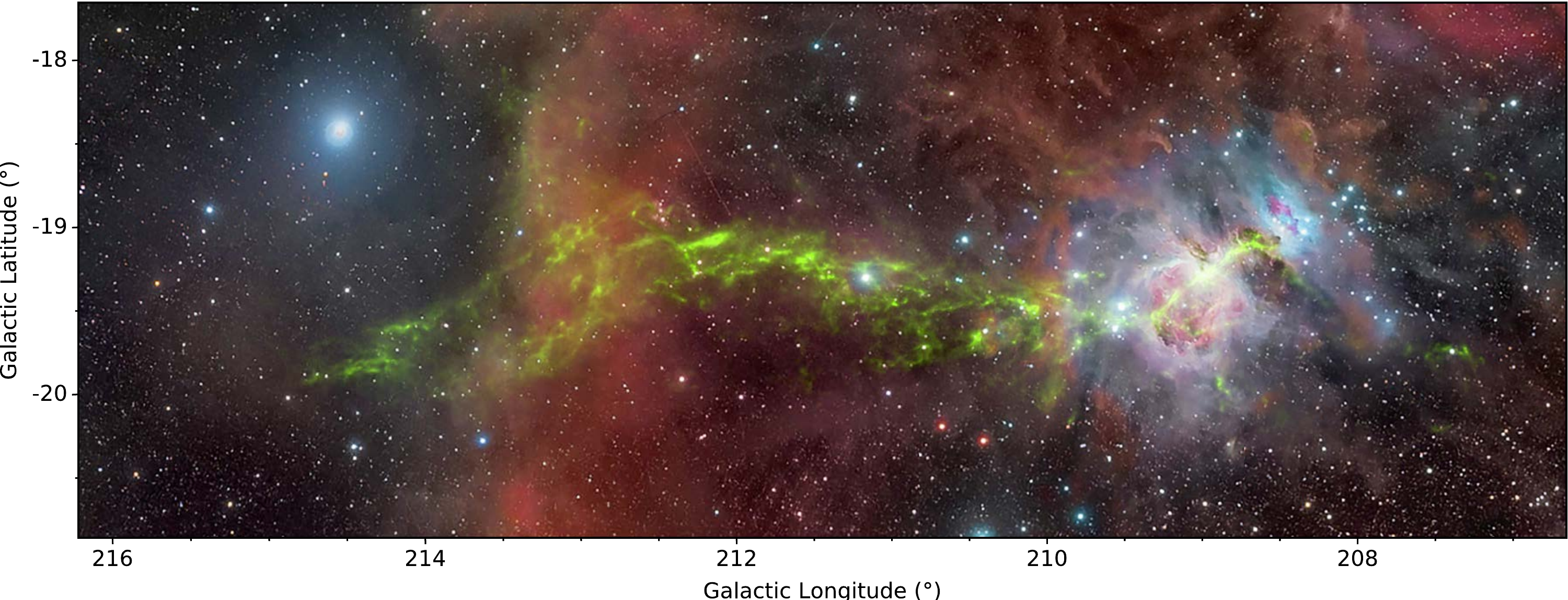}%
    \hspace{-\hsize}%
    \begin{ocg}{img:coverage_a}{img:coverage_a}{0}%
    \end{ocg}%
    \begin{ocg}{img:coverage_b}{img:coverage_b}{1}%
        \includegraphics[width=\hsize]{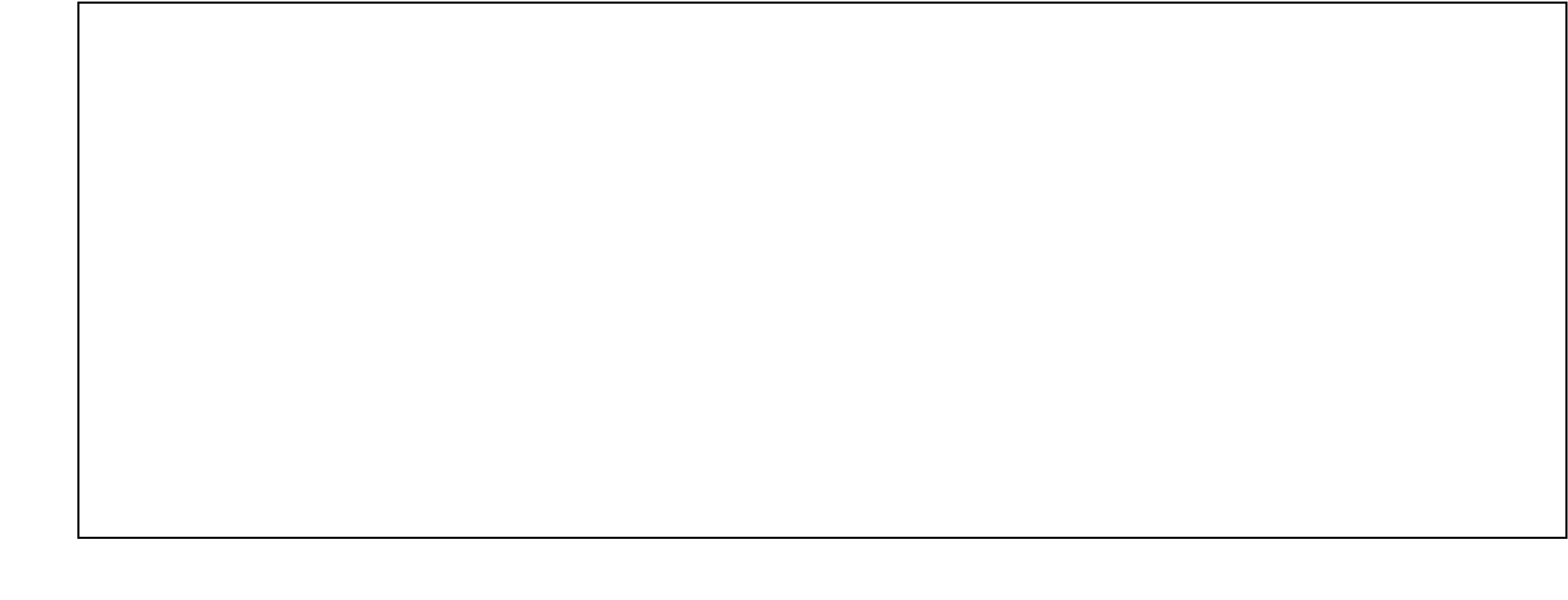}%
        \hspace{-\hsize}%
        \includegraphics[width=\hsize]{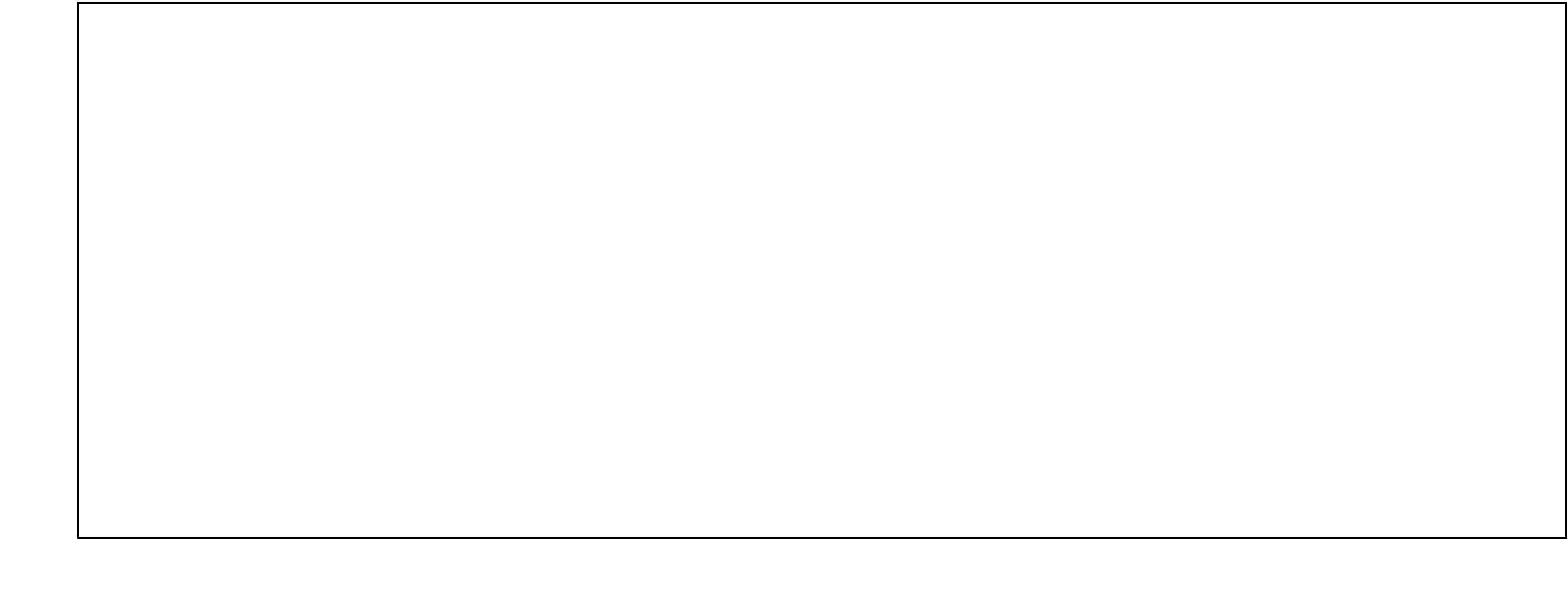}%
    \end{ocg}%
    \caption[]{Region under investigation in this work: the \object{Orion~A} molecular cloud is visualized in green with the dust optical depth map from \citet{lombardi14} blended with optical image data (\href{deepskycolors.com}{deepskycolors.com}; \textsf{\textcopyright} Roberto Bernal Andreo). The solid lines indicate the coverage of the various surveys used to investigate the properties of dust extinction in the cloud, where the IRAC contour outlines the region covered by either of the first two IRAC channels. We also make use of MIPS data, for which the covered region generally encompasses the IRAC contour. \ToggleLayer{img:coverage_b,img:coverage_a}{\protect\cdbox{toggle labels\protect\footnotemark}}}
    \label{img:coverage}
\end{figure*}

Extinction is the attenuation of electromagnetic radiation by gas and dust. The term extinction combines the effects of absorption and scattering processes which lead to a loss in the measured intensity of a light-emitting source. This effect is generally much stronger for shorter wavelengths, often obscuring sightlines associated with large amounts of dust. Including the fact that much of the absorbed energy is re-radiated thermally in the infrared, it is generally understood that these processes play a fundamental role in the determination of the radiation field of galaxies at both small and large scales. It therefore has long been recognized that measuring and understanding extinction is crucial for a large variety of astrophysical problems \citep[e.g.][]{lada94}. At the smallest scales, measuring extinction allows to constrain dust properties, such as grain sizes and element abundances \citep[e.g. based on models from][]{weingartner01, draine03}. In star-forming regions, dust extinction shields pre-stellar cores from the surrounding UV radiation. On galactic scales, dust converts a large fraction of the absorbed starlight into thermal radiation \citep{bernstein02} and at the largest scales the extra-galactic distance ladder critically depends on our understanding of the extinction law \citep[e.g.][]{nataf15}.

The shape of the extinction curve (i.e. how extinction varies with wavelength) has been the topic of many studies in the last few decades. Almost 30 years ago, \citet{cardelli89} provided the framework for the characterization of extinction curves based on the fundamental work by \citet{rieke85}, \citet{fitzpatrick86}, and \citet{fitzpatrick88}. They introduced a parameter family to describe the shape of the extinction curve and furthermore showed that an approximation is possible by using just one quantity: $R_V \equiv A_V / E_{B-V}$, with $A_V$ as the extinction in $V$ band and $E_{B-V}$ as the colour excess in $B-V$. This parameter is nowadays widely used to describe extinction characteristics in general, and in particular also to probe variations in the extinction law across large spatial scales. To this day, several key characteristics of the extinction curve remain a controversial topic. For example, at UV to optical wavelengths, the extinction curve is relatively featureless with the exception of the prominent bump at \SI{2175}{\angstrom} of which the origin still remains under debate. \citep[e.g.][]{fitzpatrick86, mathis94, bradley05}.

While in the UV and optical spectral range the extinction law is generally found to vary by a large degree for different sightlines, the NIR extinction law is widely believed to show little to no variation at all. Even though there are studies reporting different extinction parameters for the three NIR bands $J$, $H$, and $K_S$, conclusive evidence of variations depending on dust characteristics is still missing. Numerous examples on the determination of the NIR extinction law can be found in the literature. Most of these are based on the Two Micron All Sky Survey \citep[2MASS;][]{skrutskie06} where some results indicate variability of the law \citep[e.g.][]{nishiyama06, zasowski09, gosling09, wang13, nataf16}, while others do not \citep[e.g.][]{stead09, majaess16}.

Investigating an often-used power-law parametrization in the NIR ($A_{\lambda} \propto \lambda^{-\gamma}$), \citet{stead09} further complicate the situation by claiming, that the value of the power law index is critically dependent on the choice of the filter wavelength. They also continue to show that previously found variations can be explained by systematic errors. Only recently, \citet{wang14} used APOGEE \citep{eisenstein11} data to select K-type giants scattered across the galactic plane and presented convincing evidence that the NIR extinction law does not vary with colour excess. As a result, and in contrast to their previous work \citep{wang13}, this result indicates that the NIR extinction law is constant from diffuse to dense environments.

\footnotetext{This manuscript contains some figures that feature the option to switch between different layers. The implementation uses JavaScript and therefore only works within Adobe\textsuperscript{\textregistered} Acrobat\textsuperscript{\textregistered}.}   

In the mid-infrared (MIR), this picture drastically changes. Here, many different absorption features - the most prominent ones at $\sim$9.7 and \SI{{\sim}18}{\micro\metre} are caused by silicates - depend on environmental conditions. This has led to a variety of results, where some sightlines are well explained by the standard silicate-graphite model \citep{draine84, draine89, weingartner01, draine03}, while other regions (including the galactic centre) show a particularly flat distribution between 4 and \SI{8}{\micro\metre}. The origin of this flat extinction law between the NIR bands and silicate features is still not well understood \citep[e.g.][and references therein]{wang13, wang15, voshchinnikov17}.

By far the most popular targets for extinction studies in the NIR and MIR spectrum are found in the galactic plane, and in particular also near the galactic centre. Sightlines in these directions naturally provide many advantages when studying extinction in the infrared. Not only are there orders of magnitudes more sources compared to regions far away from the plane, but also distinct stellar populations can aid in the determination of the extinction law. Moreover, in recent year, the \textit{Spitzer Space Telescope} \citep{werner04}, and in particular the GLIMPSE programme \citep[][]{benjamin03, churchwell09}, have provided a large MIR database of the these regions.

There are many examples of studies inferring the MIR extinction law from \textit{Spitzer} data. Most prominently, \citet{indebetouw05} used the well-known intrinsic magnitudes of red clump giants in the galactic plane for two different sightlines to determine the absolute extinction law for the NIR $JHK_S$ and the \textit{Spitzer}/IRAC passbands. Furthermore, the regions near the galactic centre have been particularly popular to investigate the origins of the often-observed flat MIR extinction law \citep[e.g.][]{lutz96, lutz99, nishiyama09, fritz11}. Variations in the MIR extinction law have been detected in the Coalsack Nebula ($l\sim301$, $b\sim-1$) by \citet{wang13}. They find decreasing MIR extinction relative to $K_S$ band extinction $\left( A_\lambda/A_{K_S} \right)$ from diffuse to dense environments. In contrast, \citet{ascenso13} studied the dense cores B59 and FeSt 1-457 in the Pipe Nebula and do not find statistically significant evidence for a dependency of the MIR extinction law on local gas density.

In light of these often-contradicting findings, it is rather challenging to draw final conclusions about the infrared extinction law and the origin of potential variability. Most of the above mentioned studies concentrate on regions in the galactic plane and use one or another combination of photometric passbands to determine colour excess ratios as a proxy for the extinction law. In the galactic plane, however, physical conditions of the interstellar environment are rarely well-known and measured colour excesses can be influenced by multiple, physically separate, stacked regions along the line-of-sight. To reach a full understanding of the extinction law and its characteristics in the infrared, it is therefore crucial to provide observational data for regions where environmental conditions are well-known and can be factored into the result.

In the spirit of this series of papers, we will base our investigation on the \oriona giant molecular cloud \citep[Fig.~\ref{img:coverage}; for an overview see the first paper in this series and references therein:][hereafter \citetalias{meingast16}]{meingast16}. \oriona is located at a distance of \SI{414}{pc} \citep{menten07} towards the direction of the galactic anti-centre, and far below the plane. Since the region harbors many isolated star-forming events, as well as the massive young Orion Nebula Cluster (\object{ONC}), it provides a unique opportunity to study different physical conditions in a self-consistent way. Moreover, the cloud is relatively well-isolated from other star-forming regions in the vicinity. This isolation assures that any characteristics with respect to extinction are only originating from the gas and dust associated with one single molecular cloud.

In this manuscript we will investigate the extinction law in the infrared from just below \SI{1}{\micro\metre} up to about \SI{25}{\micro\metre}. Based on statistically significant variations of the colour excess ratios in several passbands, we will show that the extinction law in the infrared varies with environmental conditions. In particular we will show that the most likely explanation for these variations is the influence of the massive stars in the ONC on the surrounding interstellar medium. Their intense radiative feedback seems to impact the general dust composition and changes the overall extinction law for both NIR and MIR passbands. Comparing our findings to model predictions, we furthermore conclude, that popular dust models do not sufficiently explain the global infrared extinction law in \oriona. Based on these findings we will construct a new extinction map of the region and demonstrate that cross-calibration with \textit{Herschel} dust optical depth data can be biased by cloud substructure that is not sampled by background sources.

To arrive at these conclusions, we first introduce our data sample in Sect. \ref{sec:data} before continuing to give an in-depth overview of the applied methods in this manuscript (Sect. \ref{sec:methods}). Following this technical overview, we separate our results into two major parts. We first investigate the characteristics of the infrared extinction law in Sect. \ref{sec:law}, followed by the results with respect to our new extinction map in Sect. \ref{sec:map}. Finally, we summarize our findings in Sect. \ref{sec:summary}.

\begin{figure}
        \centering
        \resizebox{\hsize}{!}{\includegraphics[]{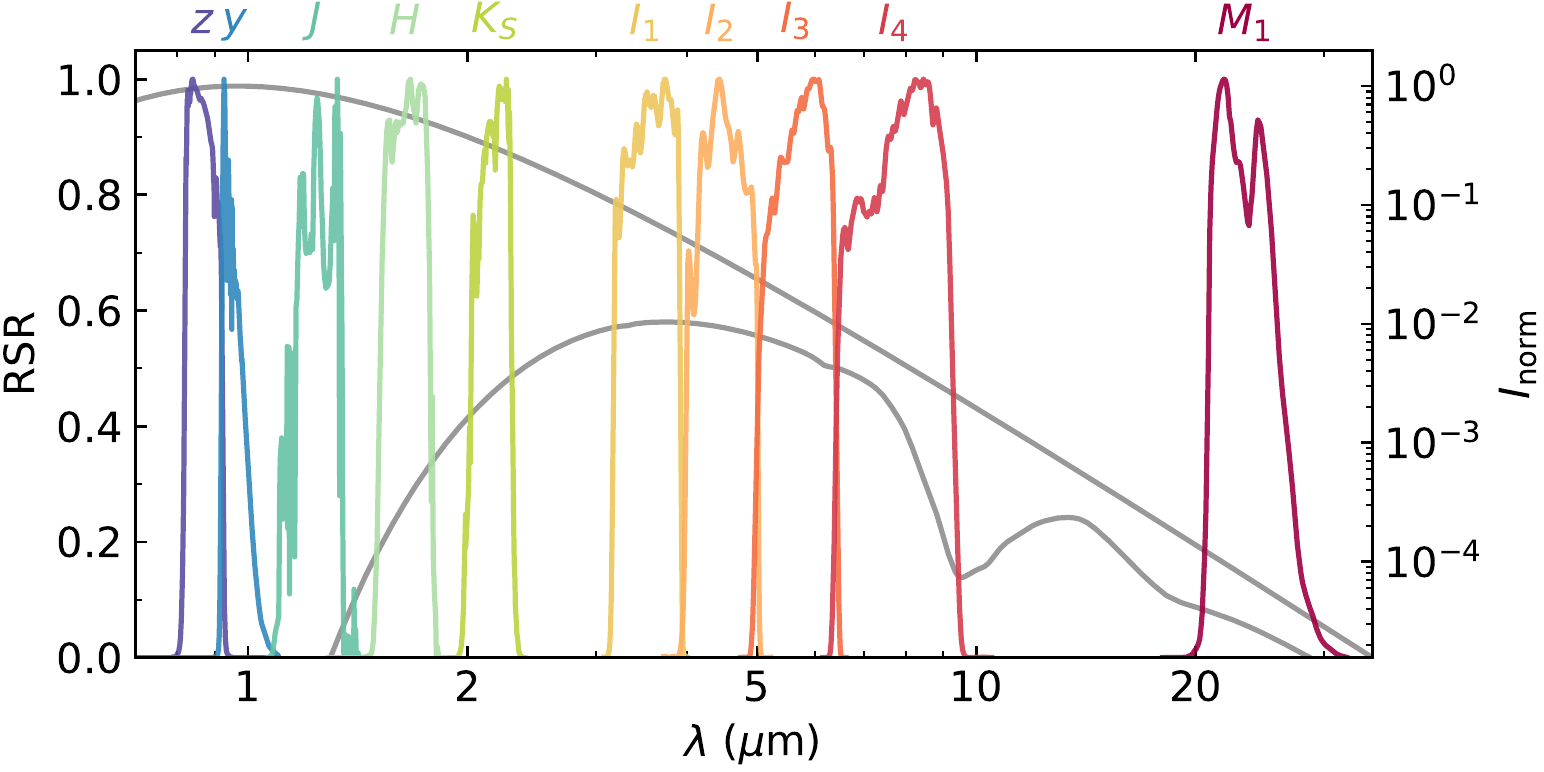}}
        \caption[]{Passbands used in our analysis cover the infrared wavelength range from just below \SI{1}{\micro\metre} to about \SI{25}{\micro\metre} utilizing the Pan-STARRS ($z$, $y$), VISTA ($J$, $H$, $K_S$; calibrated against 2MASS), and \textit{Spitzer} ($I_1$, $I_2$, $I_3$, $I_4$, $M_1$) photometric systems. The filter curves have each been normalized to their maximum transmission and their relative spectral response (RSR) is shown as a function of wavelength. The two solid grey lines are \SI{3000}{K} black bodies with a normalized intensity ($I_\mathrm{norm}$). Here, the upper line is the unmodified black body, whereas the line at the bottom shows the resulting spectral energy distribution with \SI{5}{mag} extinction in the $K_S$ band and using the $R_V = 3.1$ \citet{draine03} extinction law.}
    \label{img:passbands}
\end{figure}

\begin{figure}
        \centering
        \resizebox{\hsize}{!}{\includegraphics[]{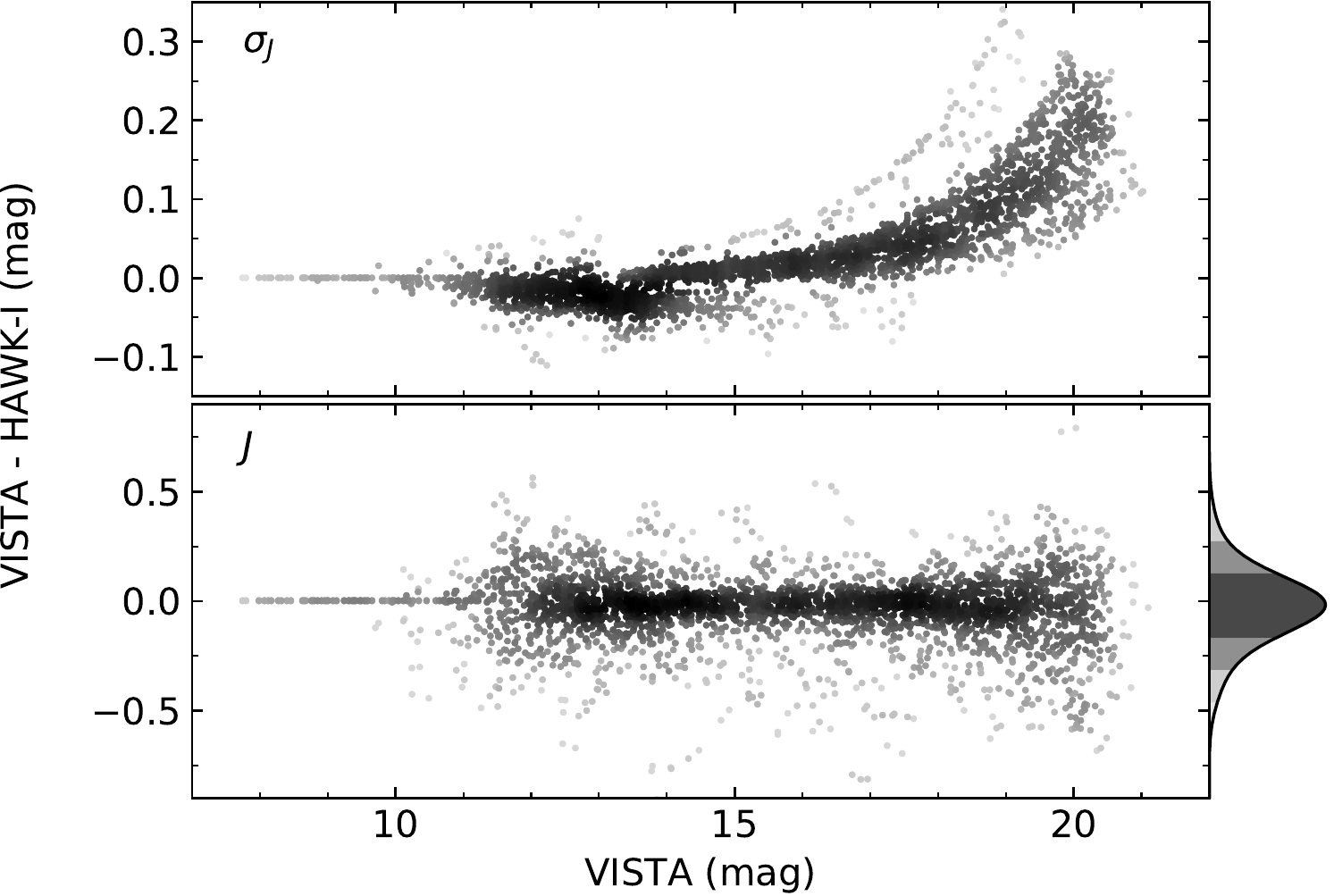}}
        \caption[]{Comparison of VISTA and HAWK-I photometry in $J$ band. The panel at the top shows the difference in photometric errors as a function of the apparent VISTA magnitude. Similarly, the bottom plot displays the difference in the measured source magnitudes. The histogram to the right shows the distribution of the magnitude difference as a kernel density. Assuming a Gaussian distribution, the differently shaded areas under the curve represent the 1, 2, and 3-$\sigma$ limits.}
    \label{img:vision_vs_hawki}
\end{figure}

\section{Data}
\label{sec:data}

This section starts with an overview of our data sources and a general description of their properties. Subsequently, we will describe how the data sets were matched, filtered, and finally combined into a master catalogue.

\subsection{Overview}

To study extinction in \oriona, we include a variety of passbands at infrared wavelengths. To this end, we combined several different sets of photometric data. The bluest bands under investigation are $z$ and $y$, supplied by the Panoramic Survey Telescope and Rapid Response System \citep[Pan-STARRS DR1;][]{Panstarrs-Surveys, Panstarrs-Calibration, Panstarrs-Database}, for which the observations cover the entire molecular cloud. The bulk of the NIR photometry is based on our work presented in \citetalias{meingast16}. In this study we introduced a large-scale imaging survey of \oriona with the Visible and Infrared Survey Telescope for Astronomy \citep[VISTA;][]{Dalton06, Emerson06} which encompasses observations in the $J$, $H$, and $K_S$ passbands. The survey covers an on-sky area of \SI{{\sim}18.3}{\deg^2} with average 90\% completeness limits of 20.4, 19.9, and \SI{19.0}{mag} in $J, H$, and $K_S$, respectively. These data are supplemented by the observations of \citet{drass16}, who used HAWK-I \citep{Kissler-Patig08} to study the \object{ONC} and its immediate surroundings. With a coverage of \SI{{\sim}0.17}{\deg^2}, they only observed a small area of the entire cloud complex. The completeness limits for these data are \SI{18}{mag} in $J$ and $H$, and \SI{17.5}{mag} $K_S$ in the innermost parts of the surveyed area (the estimated VISTA completeness limits in the same area are a few tenths of a magnitude brighter). Our analysis also covers the MIR bands as observed by \textit{Spitzer} in the four channels of the InfraRed Array Camera \citep[IRAC;][]{fazio04} and the first channel of the Multiband Imaging Photometer for \textit{Spitzer} \citep[MIPS;][]{Rieke04}. These \textit{Spitzer} data for the Orion star-forming region were provided by \citet{Megeath12} and cover most parts of \oriona which are associated with large gas column-densities (\SI{{\sim}5.86}{\deg^2} with IRAC and \SI{{\sim}11.4}{\deg^2} with MIPS). The survey borders of the $JHK_S$ and IRAC data are shown in Fig.~\ref{img:coverage}.

Since we include a total of 10 different passbands in the infrared we will use a consistent convention when referring to them throughout this manuscript. While for the bands $z$ through $K_S$ we will simply use the original filter name, the four IRAC channels will be referred to as $I_1$, $I_2$, $I_3$, and, $I_4$. Furthermore, we only have access to the first MIPS channel, which will be denoted $M_1$. In Tab. \ref{tab:sources} we list effective wavelength and width for each passband. We will also make use of the first two \textit{WISE} bands \citep{Wright2010} for calibration purposes, which will be denoted $W_1$ and $W_2$. Furthermore, we associate the passbands $z$ through $K_S$ (${\sim}0.9$ -- \SI{2.2}{\micro\metre}) with the NIR region of the electromagnetic spectrum, while the \textit{Spitzer} channels $I_1$ through $M_1$ (${\sim}3.5$ -- \SI{24}{\micro\metre}) are placed in the MIR range. An overview of our wavelength coverage is given in Fig.~\ref{img:passbands}, where the relative spectral response (RSR) curves of all passbands are displayed. The figure also shows the typical effects of extinction on a \SI{3000}{K} black body.

\begin{figure}
        \centering
        \resizebox{\hsize}{!}{\includegraphics[]{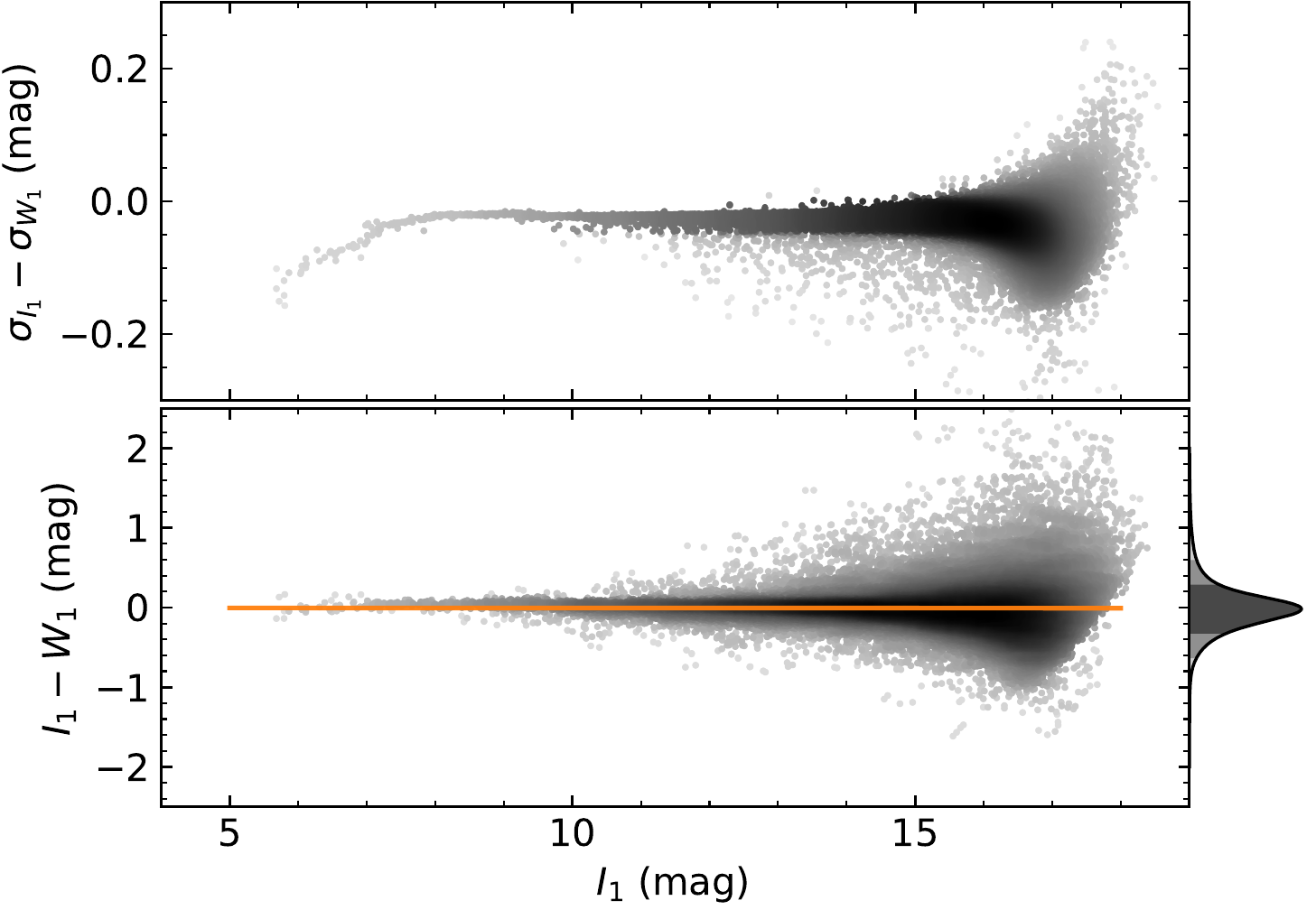}}
        \caption[]{Correlation between $W_1$ and $I_1$. The top panel shows the difference in photometric errors, while at the bottom the difference in source magnitudes is displayed. The orange line is our linear transformation between the two systems, which maps the two bands in almost a 1:1 relation. Similar to Fig.~\ref{img:vision_vs_hawki}, the histogram to the right represents the kernel density of the magnitude differences.}
    \label{img:irac1_wise1_fit}
\end{figure}

\subsection{Cross-matching and cleaning}
\label{sec:xmatch}

All catalogues were combined with TOPCAT \citep{topcat} using a 1\arcsec ~cross-matching radius where we only kept sources within the survey boundaries of the VISTA data (see Fig.~\ref{img:coverage}). All detected sources in any band of either the $JHK_S$ or the MIR catalogues were retained, independent of whether a match in another passband was found. Only for the Pan-STARRS data we required a match in any of the other catalogues due a large number of detections of nebulosity in regions with extended emission. With these criteria, the mean separation between VISTA and \textit{Spitzer} sources was 0.3\arcsec and between VISTA and Pan-STARRS sources 0.14\arcsec. For sources which were detected in multiple catalogues, we averaged their coordinates.

The $JHK_S$ catalogues from \citetalias{meingast16} and \citet{drass16} did not require photometric transformations between each other, as they are both calibrated towards the 2MASS photometric system. To minimize systematic errors, cross-matched sources between the VISTA and HAWK-I data were only retained if the magnitude difference was smaller than \SI{0.5}{mag}. For sources which are in agreement with this criterion, we calculated average magnitudes weighted by their measurement errors. For a total of 4\,340 sources in the HAWK-I catalogue, we find 3\,827 cross-matches with the VISTA catalogue (${\sim}$88\%), thus 513 new sources. About 5\% of the matched sources were then removed based on their magnitude differences. Interestingly, most of the new sources are found in a rim around the core of the ONC and are also quite easily distinguishable from the background in the VISTA image data. The fact that they were not detected in these data can be explained by the source detection procedure: While for the HAWK-I data, the entire mosaic has been searched for sources by eye, for \citetalias{meingast16} we only cleaned the very central region (11\arcmin~$\times$~11\arcmin) by hand. This also agrees well with the similar completeness limits stated above. Figure \ref{img:vision_vs_hawki} shows a comparison of all matched sources in $J$ band (the distributions for $H$ and $K_S$ look almost identical). In general, we find negligible systematic offsets between the two $JHK_S$ data sets ($\Delta$mag \SI{\sim-0.02}{mag}). The magnitude difference between the two catalogues shows a standard deviation of \SI{{\sim}0.15}{mag} across all three bands. The measurement errors, on the other hand, show discontinuities for bright sources due to different criteria on how 2MASS data was incorporated in the catalogues. As expected, for faint sources, the HAWK-I photometry features smaller errors.

\begin{table}
    \caption{Basic properties for all photometric bands used in this work, as well as corresponding source counts in our master catalogue for both the \oriona region and the control field.}
    \label{tab:sources}
    \begin{tabular*}{\linewidth}{l @{\extracolsep{\fill}} c c c c}
    \hline\hline
    Band                    & $\lambda_{\mathrm{eff}}$\tablefootmark{a} & FWHM\tablefootmark{a,b}   & \oriona   &   CF      \\
                            & \SI{}{(\micro\metre)}                     & \SI{}{(\micro\metre)}     & (\#)      & (\#)      \\
    \hline
    $z$                     & 0.87                                      & 0.10                      & 291\,259  & 54\,730   \\
    $y$                     & 0.96                                      & 0.06                      & 233\,128  & 40\,099   \\
    $J$                     & 1.24                                      & 0.22                      & 568\,116  & 84\,686   \\
    $H$                     & 1.66                                      & 0.26                      & 743\,352  & 83\,192   \\
    $K_S$                   & 2.16                                      & 0.28                      & 636\,388  & 70\,750   \\
    $I_1$\tablefootmark{c}  & 3.51                                      & 0.74                      & 192\,934  & 30\,853   \\
    $I_2$\tablefootmark{c}  & 4.44                                      & 1.01                      & 125\,951  & 25\,567   \\
    $I_3$                   & 5.63                                      & 1.39                      & 21\,126   & -         \\
    $I_4$                   & 7.59                                      & 2.83                      & 19\,324   & -         \\
    $M_1$                   & 23.21                                     & 5.32                      & 5\,914    & -         \\
    \hline
    \end{tabular*}
\tablefoot{
        \tablefoottext{a}{All parameters for the photometric passbands were obtained via the Spanish Virtual Observatory (\href{http://svo2.cab.inta-csic.es/theory/fps3/index.php}{http://svo2.cab.inta-csic.es/theory/fps3/index.php}).}
        \tablefoottext{b}{FWHM of the filter bandpass.}
        \tablefoottext{c}{\textit{Spitzer} photometry for the control field was obtained via linear transformations from \textit{WISE} (Equs. \ref{equ:sw1} and \ref{equ:sw2}).}
        }
\end{table}

As our objective was to explore characteristics of dust extinction in the molecular cloud, we are furthermore primarily interested in background sources as probes of the gas distribution. Therefore, we removed foreground stars near NGC1980, as identified by \citealp{Bouy14}, and Young Stellar Objects (YSOs) in the cloud as classified by \citet[][]{Megeath12} and Gro\ss schedl et al. (in prep.). Furthermore, upon creating the final extinction map, we found that several sources produced artefacts on the resulting map. Essentially all of these were associated with very bright stars ($K_S\lesssim$ \SI{8}{mag}), or their point spread function halos in the VISTA images and therefore were also removed from our master catalogue in an iterative procedure.

\begin{figure*}
        \centering
        \resizebox{\hsize}{!}{\includegraphics[]{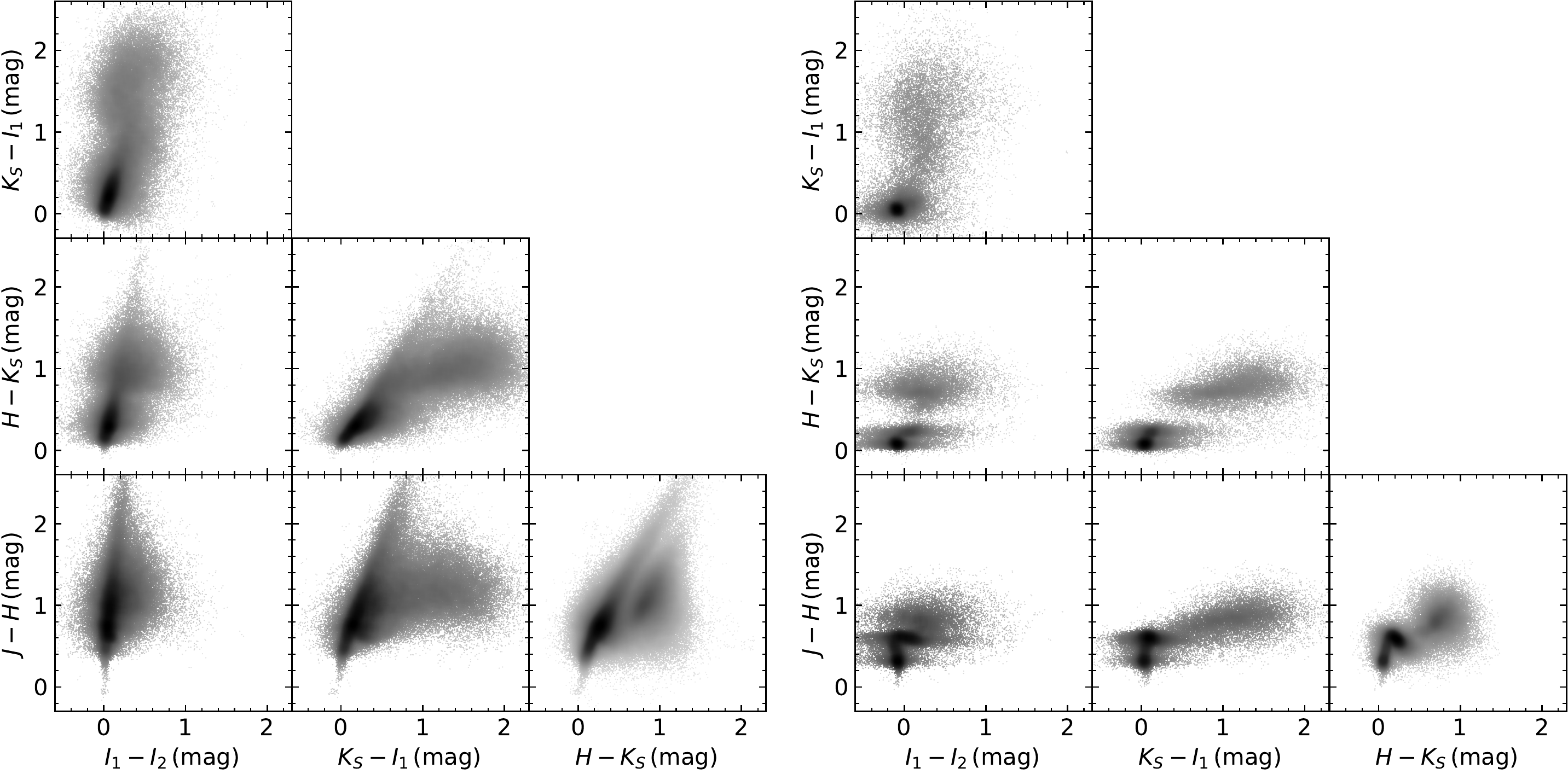}}
        \caption[]{Various colour-colour diagrams constructed from our master-catalogue. The left-hand panels show data for the \oriona region, while the panels on the right-hand side show the same combinations for the extinction free control field. Evidently, the \oriona field is heavily affected by reddening due to dust in the molecular cloud. The \textit{Spitzer} photometry for the control field was derived with a linear transformation from \textit{WISE} data and therefore shows a larger dispersion when compared to the high-sensitivity data of the \textit{Spitzer} Orion programme.}
    \label{img:combinations_scatter}
\end{figure*}

\subsection{Control field data}

For statistical comparisons of source populations we also required photometry in the same bands for an extinction-free field at approximately the same galactic latitude. $JHK_S$ photometry for a suitable control field ($l=233.3$, $b=-19.4$) is available in our data release from \citetalias{meingast16}. Pan-STARRS data for the $z$ and $y$ bands was acquired and cross-matched in the same way as for the \oriona science field. Only \textit{Spitzer} photometry was not available in this field. Fortunately, the first two channels of IRAC share very similar transmission curves when compared to the two bluest \textit{WISE} bands which enabled us to calculate a simple linear transformation between them. Due to the lower resolution of the \textit{WISE} image data, we relaxed our cross-matching radius to \SI{2}{\arcsec} between the NIR sources and \textit{WISE}, but required a detection with VISTA. In addition we required a photometric error smaller than \SI{0.5}{mag} in each band. For the fit, we obtained \textit{AllWISE} photometry \citep[][]{Cutri2013} in the corresponding bands and derived the linear transformation equations using the method later described in Sect. \ref{sec:methods:fitting}. Using the profile-fitting photometry in the catalogue (columns w1mpro, w2mpro and their associated errors), we find an almost perfect 1:1 correlation for both bands, for which the transformation equations read
\begin{align}
\label{equ:sw1}
    I_1 &= 0.999 \times W_1 + 0.011 \\ \relax
\label{equ:sw2}
    I_2 &= 1.003 \times W_2 - 0.003.
\end{align}
As an example of the well-behaved fit between the two systems, we show the correlation between $I_1$ and $W_1$ in Fig.~\ref{img:irac1_wise1_fit} (the correlation between $I_2$ and $W_2$ looks almost identical).

\subsection{Master catalogue}

With the criteria listed above we constructed our final master catalogues for both the \oriona, as well as the control field. Number counts and basic passband properties of all bands under consideration are tabulated in Table \ref{tab:sources}. Furthermore, a selection of colour-colour combinations is displayed in Fig.~\ref{img:combinations_scatter} for both the \oriona region (left-hand panels) and the control field (right-hand panels). Clearly, the \oriona field is affected by a significant amount of reddening when compared to the extinction-free control field. In almost all diagrams a clear distinction between extragalactic sources and main sequence stars is visible (e.g. $J-H\sim1$, $H-K_S\sim0.8$, $K_S-I_1\sim1.2$). Another interesting feature in the \oriona diagrams is the clearly discernible changing extinction vector: the slope in the diagrams gets significantly steeper from $H-K_S$ to $I_1-I_2$ indicating decreasing extinction differences in the two bands constituting each colour. For example, in the diagrams including $I_1-I_2$, the extinction vector is almost vertical.

A closer look at the control field diagrams, including the transformed \textit{Spitzer} photometry, reveals a non-negligible amount of additional noise compared to the original \textit{Spitzer} \oriona photometry. Most prominently, the stellar main sequence at $I_1 - I_2 \sim 0$ is significantly broader in the control field. Quantifying this additional noise component is difficult since the control field sample is not affected by extinction, whereas the \textit{Spitzer} observations mostly cover areas associated with the molecular cloud. Limiting both samples to $J-H<$~\SI{0.5}{mag} and applying various magnitude cuts from $I_1=15$ to \SI{17}{mag} effectively removes many sources affected by extinction, as well as galaxies. Comparing the resulting distributions for the $I_1 - I_2$ colour, we find an about 50\% broader stellar main sequence for the control field data and directly interpret this increase as an estimate for the additional noise. This presents a problem for statistical comparisons and will examine this issue further in Sect. \ref{sec:map} where these diagrams are used to construct an extinction map. For the remainder of this article, all further analysis is based on this master-catalogue and throughout the individual sections we will refer to various subsets on multiple occasions.

\section{Methods}
\label{sec:methods}

In this section we will give an in-depth overview of all methods used in this manuscripts. In particular, this includes mathematical background on how we fit observed noisy data with linear relations (Sect. \ref{sec:methods:fitting}), how to derive an infrared extinction law (Sect. \ref{sec:methods:law}), and how to measure extinction for point-sources (Sect. \ref{sec:methods:pntsrc}). Finally, we will also describe how to use these point-source measurements to construct the smooth gas distribution in the ISM in the plane of the sky with extinction mapping procedures (Sect. \ref{sec:methods:mapping}).

\subsection{Linear fitting}
\label{sec:methods:fitting}

Many results derived in this manuscript depend on reliable linear fitting procedures. Moreover, our interpretations and analysis also depend on the statistical significance of the resulting fit. For this purpose, we develop in this section a robust Bayesian framework which takes errors in both dimensions into account and ultimately enables us to derive reliable model parameters. The resulting framework is then employed in combination with a Markov chain Monte Carlo (MCMC) ensemble sampler. Specifically, we define a likelihood function which can be numerically optimized and also can, in principle, take any form, but must be an appropriate choice for the given data characteristics. The MCMC algorithm derives suitable model parameters by sampling the parameter space in a so-called chain. Individual positions in the chain are chosen based on the used samplers \citep[e.g. Metropolis;][]{mackay03} and at each position in the chain, the posterior probability is evaluated. The probabilities at each step constitute a trace and these traces allow statistical evaluations for each parameter. Here, we always choose to use the mean and the standard deviation of the traces to describe the model parameters and its uncertainties. For reasons of simplicity and reproducibility we chose the emcee\footnote{Description, documentation, and source code are available at \href{http://dan.iel.fm/emcee/current/}{http://dan.iel.fm/emcee/current/}} MCMC ensemble sampler \citep{emcee, hogg10}. This implementation allows to optimize model parameters with respect to a given pre-defined model likelihood. Details on how emcee constructs the chains are given in \citet{emcee}.

The MCMC approach requires to define a model for our data and an associated likelihood. We start by defining a simple linear relation between generic values on the abscissa and the ordinate in a Euclidean space, which is given by
\begin{equation}
    \label{equ:model}
    y (x ~ | ~ \alpha, \beta) = \alpha + \beta x
\end{equation}
where $\alpha$ is the intercept, and $\beta$ the slope\footnote{Throughout this manuscript we will use the variable $\beta$ to generally refer to the slopes of linear relations. When the slope is used to derive a specific quantity $\beta$ will be labelled with an appropriate subscript.}. Furthermore, we define our generic observed dataset as
\begin{equation}
    \label{equ:dataset}
    X = \{ x_i, y_i \}
\end{equation}
where the $x_i$ and $y_i$ are the measurements with corresponding errors $\sigma_{x,i}$ and $\sigma_{y,i}$. We also define the unknown true values $\bar{x_i}$ and $\bar{y_i}$, which are connected to the above set via
\begin{align}
    x_i &= \bar{x_i} + \epsilon_{x_i} \\
    y_i &= \bar{y_i} + \epsilon_{y_i}
.\end{align}
Here we a adopt a normal distribution for the errors $\epsilon$ in both coordinates such that
\begin{align}
    \epsilon_{x_i} &= N(0, \sigma_{x_i}^2) \\
    \epsilon_{y_i} &= N(0, \sigma_{y_i}^2)
.\end{align}
Given these assumptions and our data set, we can derive a model likelihood, for which the dispersions in $x$ and $y$ along the linear relation are described by normal distributions, which in turn are characterized by the measurement errors. Thus, the probability for each data point, given $\alpha$ and $\beta$ defining a model as in Equ. \ref{equ:model}, is
\begin{align}
\begin{split}
\label{equ:prob_single}
P(x_i, y_i ~ | ~ \alpha, \beta, \bar{x_i}, \sigma_{x,i}, \sigma_{y,i}) =& \frac{1}{\sqrt{2\pi} \sigma_{y_i}} \mathrm{exp} \left[ -\frac{\left( y_i - \alpha - \beta \bar{x_i} \right)^2}{2\sigma_{y_i}^2} \right] \times \\
& \frac{1}{\sqrt{2\pi} \sigma_{x_i}} \mathrm{exp} \left[ -\frac{\left( x_i - \bar{x_i} \right)^2}{2\sigma_{x_i}^2} \right]
\end{split}
,\end{align}
where we already take the linear relation, described by the intercept $\alpha$ and slope $\beta$, into account to eliminate the true values $\bar{y_i}$. We note here that Equ. \ref{equ:prob_single} assumes uncorrelated errors for $x_i$ and $y_i$. Strictly speaking, in many of our applications of the linear fitting technique in this manuscript, we do have correlated errors in both dimensions (e.g. in colour-colour diagrams where a given passband occurs in both colours). However, a generalization of the probability and resulting likelihood to account for correlated errors is far from trivial and is beyond the scope of this paper. Given the excellent results of the technique, we do not expect the assumption of uncorrelated errors to have a major impact on our results.

The above given definition in Equ. \ref{equ:prob_single}, still includes the unknown true values of $x_i$, $\bar{x_i}$. For our application we therefore need to find a way to eliminate them as well. In the following lines we will derive our final model likelihood along which we will also deal with this issue. The full posterior probability distribution of the model parameters $\alpha$ and $\beta$, given the set of measurements $X$ can be rewritten to include the true values $\bar{x_i}$.
\begin{equation}
\label{equ:posterior1}
    P(\alpha, \beta ~ | ~ X, \sigma_{x,i}, \sigma_{y,i}) = \int P(\alpha, \beta, \bar{x_i} ~ | ~ X, \sigma_{x,i}, \sigma_{y,i}) \, \mathrm{d}\bar{x_i}
\end{equation}
Following Bayes' theorem, the term in the integral can be written as
\begin{equation}
\label{equ:posterior2}
    P(\alpha, \beta, \bar{x_i} ~ | ~ X, \sigma_{x,i}, \sigma_{y,i}) \propto P(X ~ | ~ \alpha, \beta, \sigma_{x,i}, \sigma_{y,i}, \bar{x_i})  \; P(\alpha, \beta, \bar{x_i})
\end{equation}
The right-hand side of this relation splits up into the likelihood function $P(X ~ | ~ \alpha, \beta, \sigma_{x,i}, \sigma_{y,i}, \bar{x_i})$ and the so-called prior function $P(\alpha, \beta, \bar{x_i})$. For simplicity, we assume a flat prior on $\bar{x_i}$, i.e. $P(\bar{x_i})=1$. Also, the normalization (the denominator in Bayes' theorem) can be ignored in our case since it is a constant and would only become important if we attempted to compare different model definitions. Thus, Equ. \ref{equ:posterior1} becomes
\begin{equation}
\label{equ:posterior3}
    P(\alpha, \beta ~ | ~ X, \sigma_{x,i}, \sigma_{y,i}) \propto P(\alpha, \beta) \!\int\! P(X ~ | ~ \alpha, \beta, \sigma_{x,i}, \sigma_{y,i}, \bar{x_i}) \, \mathrm{d}\bar{x_i}
.\end{equation}
The likelihood function for the data ensemble $X$, given by $P(X ~ | ~ \alpha, \beta, \sigma_{x,i}, \sigma_{y,i}, \bar{x_i})$, can be constructed by multiplying over all $i$
\begin{equation}
P(X ~ | ~ \alpha, \beta, \sigma_{x,i}, \sigma_{y,i}, \bar{x_i}) = \prod_{i} P \left( x_i, y_i ~ | ~ \alpha, \beta, \bar{x_i}, \sigma_{x,i}, \sigma_{y,i} \right)
,\end{equation}
which, together with Equ. \ref{equ:prob_single} can be inserted into Equ. \ref{equ:posterior3}, giving
\begin{align}
\begin{split}
\label{equ:posterior4}
   P(\alpha, \beta ~ | ~ X, \sigma_{x,i}, \sigma_{y,i}) \propto & P(\alpha, \beta) ~ \times \\ & \prod_{i} \int\! \frac{1}{\sqrt{2\pi} \sigma_{y_i}} \mathrm{exp} \left[ -\frac{\left( y_i - \alpha - \beta \bar{x_i} \right)^2}{2\sigma_{y_i}^2} \right] \times \\
   & \frac{1}{\sqrt{2\pi} \sigma_{x_i}} \mathrm{exp} \left[ -\frac{\left( x_i - \bar{x_i} \right)^2}{2\sigma_{x_i}^2} \right] \, \mathrm{d}\bar{x_i}
\end{split}
.\end{align}
This relation contains an integral over the product of two Gaussians, that is, a convolution, and thus can be rewritten as a single normal distribution. Furthermore, by rearranging the argument of the exponential term in the first Gaussian function, we are also able to eliminate $\bar{x_i}$ during this process. The integral in the above equation then evaluates to
\begin{equation}
    \int P(X) \, \mathrm{d}\bar{x_i} = \frac{1}{\sqrt{2 \pi (\sigma_{y_i}^2 + \beta^2 \sigma_{x_i}^2)}} \mathrm{exp}\left[ -\frac{(y_i - \alpha - \beta x_i)^2}{2 (\sigma_{y_i}^2 + \beta^2 \sigma_{x_i}^2)} \right]
.\end{equation}
Defining
\begin{equation}
    \xi^2_{i} = \sigma_{y,i}^2 + \beta^2 \sigma_{x,i}^2
\end{equation}
allows to rewrite our likelihood function which finally reads in logarithmic form
\begin{equation}
\label{equ:likelihood}
    \ln P \left(X ~ | ~ \alpha, \beta, \xi_{i} \right) \propto -\frac{1}{2} \sum_i \left[ \ln \left( 2\pi \xi_{i}^2 \right) + \frac{\left( y_i - \alpha - \beta x_i \right)^2}{\xi_{i}^2} \right]
.\end{equation}
This likelihood allows to efficiently derive optimized linear fitting parameters, taking measurement errors in both dimensions into account. Interestingly, \citet{clutton67} also investigated this issue and arrive at the same final likelihood definition, using a more traditional approach.

For the posterior probability defined in Equ. \ref{equ:posterior2} we also need the prior function $P(\alpha, \beta)$. Here, we chose a uniform prior for our application and write the logarithmic prior for the intercept $\alpha$ and the slope $\beta$ in the form of a piecewise function:
\begin{equation}
\label{equ:prior}
\ln P(\alpha, \beta) =
  \begin{cases}
    0       &   \alpha_{\mathrm{low}} \leq \alpha \leq \alpha_{\mathrm{high}} ~ \land ~ \beta_{\mathrm{low}} \leq \beta \leq \beta_{\mathrm{high}} \\
    -\infty &   \mathrm{otherwise}
  \end{cases}
,\end{equation}
where $\alpha_{\mathrm{low}}$, $\alpha_{\mathrm{high}}$, $\beta_{\mathrm{low}}$, and $\beta_{\mathrm{high}}$ are reasonable choices for the given application. For instance, when fitting colour-colour diagrams, we know that the intercept must be close to 0. We will, however, be very conservative with the choice of our priors, which mostly are needed for faster convergence of the numerical optimization process. Finally, with the prior given in Equ. \ref{equ:prior} and the likelihood given in Equ. \ref{equ:likelihood} we can calculate the posterior probabilities of the model parameters $\alpha$ and $\beta$ from Equ.~\ref{equ:posterior3} as
\begin{equation}
    \ln P(\alpha, \beta ~ | ~ X, \xi_i) \propto \ln P(X ~ | ~ \alpha, \beta, \xi_i) + \ln P(\alpha, \beta)
\end{equation}

We want to close this section on our linear fitting algorithm by acknowledging that we are fully aware that our choice of a Gaussian distribution is not always strictly true. However, for most problems it is extremely challenging to define accurate likelihood functions. For instance, when fitting colour-colour diagrams, the intrinsic distribution of stellar colours and the dependence of this distribution on extinction (faint low-mass stars have different colours than bright high-mass stars) produce a highly complex spread in colours along the extinction vector, which is very difficult to model without prior information on the intrinsic source characteristics. Therefore, choosing a Gaussian likelihood presents a very good compromise between the actual distribution and neglecting these errors in the first place, as is usually done when facing such problems.

\subsection{Measuring extinction}
\label{sec:methods:extinction}

In the following sections we discuss our procedures to derive an extinction law from photometric data, to extinction for single sources, and to construct the smooth column-density distribution from these measurements.

\subsubsection{Deriving an extinction law}
\label{sec:methods:law}

Many previous studies have derived extinction laws for a variety of different sightlines. In general, methods relying on spectroscopic data have proven to be very effective for determining the line-of-sight extinction as a function of wavelength for individual sources. For these, the underlying procedure relies on comparing spectra of the target of interest to an extinction-free source. On the other hand, the availability of photometric surveys, covering large spectral and spatial ranges, has also led to the development of techniques to derive extinction laws for the various available photometric passbands.

In general, these methods can be split into absolute determinations of the extinction law with respect to wavelength, and relative measurements. In the case of absolute determinations, a direct calibrator is required. Noteworthy examples of such calibrations can be found in \citet{indebetouw05}, \citet{schlafly10}, and \citet{schlafly11}. \citet{indebetouw05} parametrized the apparent magnitudes of red clump giants to determine the interstellar reddening as a function of distance. \citet{schlafly10} measured the main-sequence turnoff at the blue edge for various SDSS colours and then fitted this ``blue tip'' with a reddening-dependent relation, whereas \citet{schlafly11} compared measured SDSS colours to model-predicted source colours.

In contrast, studies determining only a relative dependence of the extinction law require a zero-point to calibrate their results \citep[e.g.][]{flaherty07, ascenso13}. In our case, we will also only determine a relative extinction dependence, mainly because it is very challenging to establish an absolute law for the specific case of the \oriona molecular cloud based on photometry. One can easily imagine that techniques relying on a distinct population of red giant stars can not be applied here due to the location of the cloud near the galactic anti-centre and ${\sim}20\degr$ below the plane. Also, our limited field-of-view which, to a large extent, is heavily affected by dust-extinction, makes statistical determinations of features in colour space (e.g. the blue edge) unreliable without prior assumptions on the dust distribution. Another reason for our choice of a relative calibration is that the interpretation of our findings does not depend on a direct calibration.

Here, we will derive a wavelength-dependent extinction law for a variety of photometric passbands based on the source distribution in a set of colour-colour diagrams. The basis for this method lies in the definition of a source's colour excess
\begin{align}
\label{equ:excess1}
E_{m_1 - m_2}   &= (m_1 - m_2) - (m_1 - m_2)_0   \\
\label{equ:excess2}
                &= (m_1 - m_{1,0}) - (m_2 - m_{2,0}) = A_{m_1} - A_{m_2}
,\end{align}
where $E_{m_1 - m_2}$ is the notation for the colour excess measured in the passbands $m_1$ and $m_2$ (e.g. $J$ and $H$). This excess is simply the difference between the measured apparent colour ($m_1 - m_2$) and the intrinsic colour $(m_1 - m_2)_0$ of the same source, or the difference in extinction in the two passbands ($A_{m_1} - A_{m_2}$). Using two colour excesses in differently combined passbands the colour excess ratio $\beta_{m_{\lambda}}$ is defined as
\begin{equation}
    \label{equ:beta}
    \beta_{m_{\lambda}} = \frac{E_{J - m_{\lambda}}}{E_{J - K_S}} = \frac{A_J - A_{m_{\lambda}}}{A_J - A_{K_S}}
,\end{equation}
and represents the slope in the diagram plotting the two parameters against each other. Here, $m_{\lambda}$ denotes one of the several passbands we investigate in this manuscript and $A_{m_{\lambda}}$ refers to the extinction in this particular band. Furthermore, for all applications we will use $J - K_S$ as the baseline of this parameter. This choice offers both a large range in measured colours ($\Delta(J - K_S) > 8$ for many areas) and superior number statistics.

Upon extending the total extinction ratio $A_{m_{\lambda}} / A_J$ and substituting with a rearranged Equ. \ref{equ:beta} we find
\begin{equation}
    \label{equ:a_lambda}
    \frac{A_{m_{\lambda}}}{A_J} = \frac{A_J - (A_J - A_{m_{\lambda}})}{A_J} = \beta_{m_{\lambda}} \left( \frac{A_{K_S}}{A_J} - 1 \right) + 1
.\end{equation}
Using this equation, we see that the total extinction ratio for any passband can be determined by measuring the slope $\beta_{m_{\lambda}}$ and assuming a value of $A_{K_S} / A_J$. For the remainder of this manuscript we adopt $A_J / A_{K_S} = 2.5 \pm 0.15$ $ \left( A_{K_S} / A_{J} = 0.4 \pm 0.024 \right)$ as published by \citet{indebetouw05}. However, since this conversion necessarily includes further systematic errors, we prefer to make any comparisons directly with the fitted colour excess ratio.

A closer examination of the equations \ref{equ:excess1}, \ref{equ:excess2}, and \ref{equ:beta} reveals that the determination of $\beta_{m_{\lambda}}$ requires knowledge about the intrinsic source colours. It is, however, possible to bypass this requirement. Ignoring differences across the filter bandpasses due to varying spectral energy distributions, all sources in a colour-colour diagram are pushed in the same direction along the extinction vector in a diagram plotting the apparent colours of the sources. Thus, the slope in the distribution of apparent source colours equals the colour excess ratio in Equ. \ref{equ:beta}.

One caveat of this assumption is tied to the intrinsic source colour distribution in colour-colour diagrams. That is to say that the intrinsic colour distribution of the stellar main sequence is not parallel to the extinction vector which can introduce a bias in the measured slope. \citet{ascenso12} developed a technique to fit colour-colour diagrams taking this effect into consideration. However, preliminary tests with our data showed that the application of this method is unreliable in our case because the intrinsic source colour distribution rapidly changes for fields with variable extinction. In other words, for large amounts of reddening, only intrinsically bright background sources are visible through the cloud, while for regions with low extinction, the full stellar main sequence along with a set of galaxies are sampled in the diagram. For example, when searching for a spatially variable extinction law, this difference in the sampled source population introduces a bias in the fitting results, because the intrinsic colour distribution will be different for the individual fields. In contrast, when dealing with a spatially limited field with a well defined background source population, this method minimizes the bias introduced by the distribution of intrinsic source colours \citep[for an application, see][]{ascenso13}.

In our case, the influence of the intrinsic source colour distribution is minimal in the first place, since the fields we investigate show a very large range of apparent source colours. Typically we find $\Delta(J - K_S) >$ \SI{8}{mag} for our fields, whereas changes in intrinsic stellar colours are mostly limited to $\Delta(J - K_S) <$ \SI{1}{mag}. Thus, the slope difference between intrinsic source colours and the distribution generated by pushing sources along the extinction vector only leads to a broadening of the source distribution along the extinction vector. Using our linear model defined in Sect. \ref{sec:methods:fitting} this broadening will be taken into account when fitting the various colour-colour diagrams by introducing a systematic noise term which will later be discussed in the relevant sections.

\subsubsection{Estimating line-of-sight extinction with photometry}
\label{sec:methods:pntsrc}

Over the last decades a number of methods have been developed to derive extinction from point-source measurements including the well-established techniques relying on measured stellar colours \citep[e.g.,][]{lada94, alves98, lombardi01, majewski11, juvela16}. In this work we use the recently published \pnicer method \citep{meingast17} to derive the extinction towards point sources using their measured colours with a given extinction law. Briefly summarized, \pnicer builds on and extends some of the above mentioned methods. The routine applies machine learning techniques to photometric measurements of extincted sources, in combination with reddening-free control field data, to determine the extinction along the line of sight. In particular, \pnicer fits Gaussian Mixture Models along the extinction vector in arbitrary numbers of dimensions to derive probability densities describing the extinction for each source. This process also effectively bypasses often-used simplifications of the reddening-free source distribution. In this way, \pnicer circumvents many issues related to the more traditional methods mentioned above when highly sensitive photometric data are used. For example, the bias introduced by comparing the source population of highly extincted regions with extinction-free control field data can be eliminated by choosing a large number of dimensions (see Fig. 4 in \citealp{meingast17}). \pnicer does neither require any prior knowledge of the projected gas density distribution, nor does it rely on any model predictions. The method is strictly data-driven and outperforms the above mentioned techniques when the observed source population includes a significant amount of late-type stars and galaxies which can introduce a complex pattern in the colour distribution of the observed sources. Furthermore, \pnicer also is capable of reproducing results derived with the established methods for more simple setups (e.g. when using 2MASS data) and is therefore applicable in a multitude of setups.

\subsubsection{Extinction mapping}
\label{sec:methods:mapping}

As with determining the line-of-sight extinction towards single sources, there are also a number of mapping techniques available to construct the smooth projected gas surface density from photometric measurements. Traditionally, methods relying on the number density of sources have been used as a proxy for the density-distribution of gas in the interstellar medium \citep[e.g.,][]{bok73, cambresy99, dobashi05} and these continue to be useful to this day \citep[e.g.][]{beitia17}. Extinction measurements based on stellar colours have also been widely used in the past. Based on the work of \citet{lada94}, \citet{lombardi01} introduced the multi-band technique \nicer which uses a weighted average smoothing process to calculate an extinction map \citep[see also][]{alves01}. The pixel values in such extinction maps are weighted averages derived from line-of-sight ``pencil-beam'' measurements towards single sources. To combine this method with \pnicer, we convert the derived extinction probability densities to discretized values with the expected value of the distribution (see Sect. 3.3 in \citealp{meingast17}).

Extinction mapping techniques based on discrete measurements of line-of-sight extinctions all suffer from unresolved substructure in the cloud: A given set of stars in a defined region (e.g. a pixel of an extinction map) only provides a limited number of samples of the (presumably smooth) column-density distribution of a molecular cloud. Moreover, the measured background sources have a higher probability to sample the low-column density parts of a cloud because larger extinctions push more background sources beyond the sensitivity limit of the observations. To counter this effect, \citet{cambresy02} introduced a method which combines colour-based reddening calculations with star counts. Later, \citet{dobashi08} introduced a percentile-based method to minimize this bias and finally \citet{lombardi09} published the so-called \nicest method which has already found many applications in recent years \citep[e.g.][]{lombardi11, alves14}. \nicest adds an additional weighting factor - $10^{\alpha_{\textsc{Nicest}} k_{m_\lambda} A_{m_{\lambda}}}$ - to the spatial smoothing process which attempts to correct for the unresolved cloud substructure. Here, $\alpha_{\textsc{Nicest}}$ refers to the slope in the expected number counts and the factor $k_{m_\lambda}$ represents the extinction law in the passband $m_\lambda$. When combining higher numbers of dimensions, however, these coefficients are not straight-forward to determine. We will therefore derive the best-fitting value in an iterative procedure.

\section{Infrared extinction law in \oriona}
\label{sec:law}

In this section we present, discuss, and interpret our findings with respect to the infrared extinction law in \oriona. Firstly, we will derive an average extinction law for the entire survey region. Subsequently, we continue to investigate spatial variations in our sample, by separately exploring the effects of the local environment, as well as a potential dependence on gas (column-)density.

\begin{figure}
        \centering
        \resizebox{\hsize}{!}{\includegraphics[]{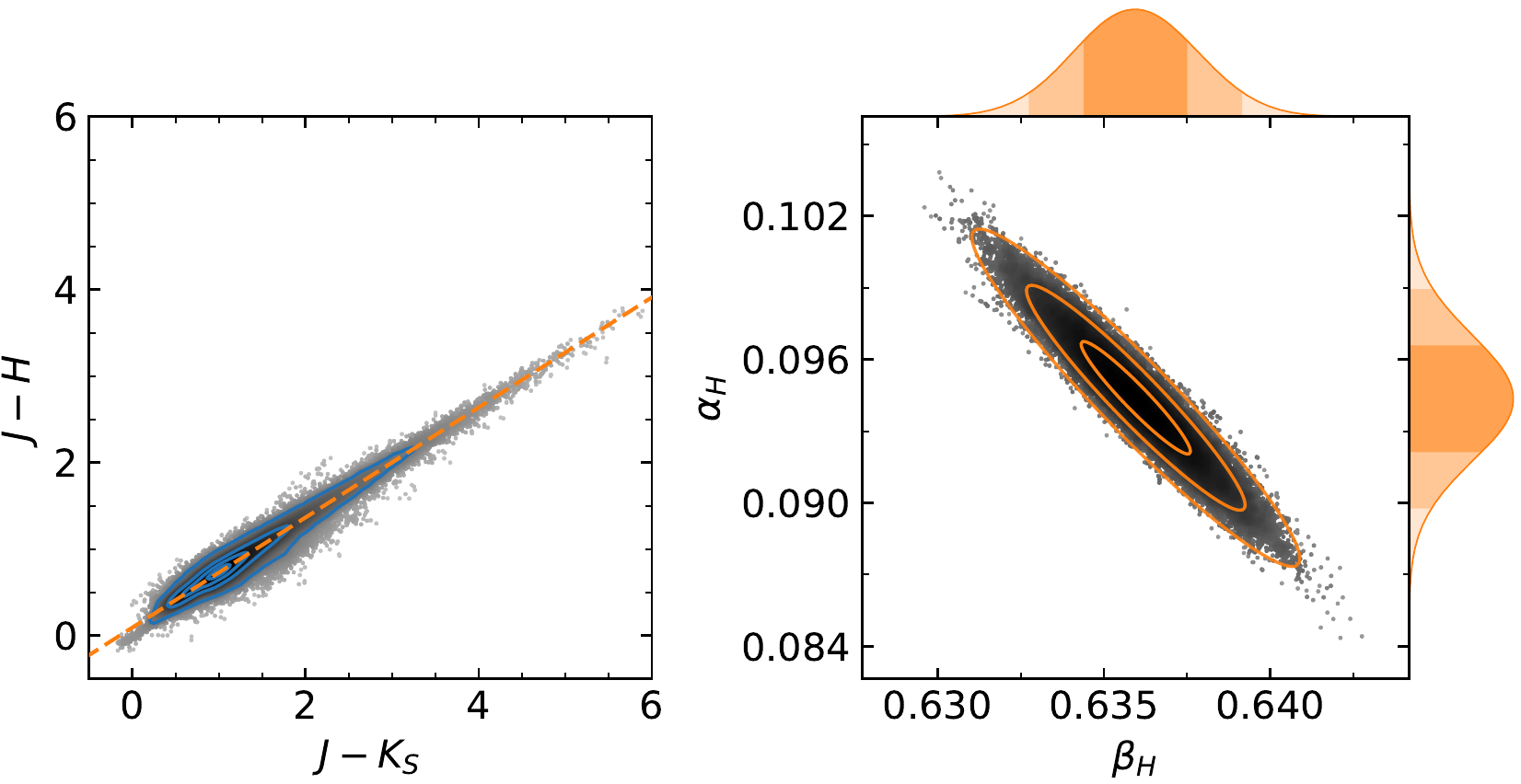}}
        \caption[]{Fitting results for the colour excess ratio in $H$ band. The panel on the left-hand side shows the filtered source distribution in a greyscale scatter plot, where the greyscale is proportional to the number density in this parameter space. The thin blue contours refer to 0.5, 10, 40, and \SI{80}{\%} of the maximum source density. Also plotted is the resulting fit as the orange dashed line. The panel on the right-hand side displays the posterior probability distribution for the intercept $\alpha_H$ and the slope $\beta_H$ (the colour excess ratio). The ellipses are the 1-, 2-, and 3-sigma covariances of the distribution. Projected to the top and to the right are normalized kernel densities of the resulting distribution for each parameter. The differently shaded filled areas under the curve highlight the 1-, 2-, and 3-$\sigma$ ranges.}
    \label{img:total_H}
\end{figure}

\subsection{Fitting colour-colour diagrams}
\label{sec:fit_ccds}

As outlined in Sect. \ref{sec:methods:extinction}, we use a series of colour-colour diagrams with various passband combinations to determine colour excess ratios as a proxy for the extinction law. By investigating Fig.~\ref{img:combinations_scatter} in more detail, it becomes apparent that all such combinations show a rather complex distribution in the multivariate colour spaces. Since we are interested only in the effects of reddening, ideally only a well defined subset with sources of similar intrinsic colours should be used to fit the relation \citep[e.g.][]{wang14}. As we do not have thorough source classifications for the entire sample, we start by reducing the amount of contamination with various filtering steps. To this end, subsets for each band have been defined to reduce the dispersion, resulting from intrinsic source colours, as well as the contamination by galaxies, nebulosity, and other spurious detections. Each band was treated separately with the following conditions:
\begin{itemize}
    \item Only sources with photometric errors smaller than \SI{10}{\%} were considered.
    \item For the Pan-STARRS bands, each source had to have at least ten independent detections ($nz$, $ny$ > 10). This criterion significantly reduced the amount of nebulous detections.
    \item For \textit{Spitzer} data we only considered sources brighter than (14, 13, 13) mag in $(I_1, I_2, I_3)$. The main purpose here was also to reduce the amount of false detections (for $I_4$, the error criterion above was enough to create a clean subset).
    \item To effectively remove extended sources, which in general have significantly different intrinsic colours compared to stars, we used the flags from the catalogue in \citetalias{meingast16} (Class\_cog = 1 $\land$ Class\_sex > 0.9).
    \item In addition to the criteria listed above, a few dozen remaining contaminating sources were removed by hand.
\end{itemize}
Applying the above criteria to our master-catalogue resulted in a remarkably clean selection of bright main sequence stars. Nevertheless, this reduced set still contained sources of different intrinsic colours. Intrinsic colours of stars show natural broad sequences, which are, in general, not parallel to the extinction vector. Thus, an additional systematic noise term is present in these data and as a consequence, the dispersion in both axes along the extinction vector is not fully described by the photometric measurement errors anymore. Due to these characteristics, extincted sources along the reddening vector are affected by a systematic broadening in addition to their photometric errors. For accurate linear fitting, however, this term needs to be accounted for.

\begin{figure}
        \centering
        \resizebox{\hsize}{!}{\includegraphics[]{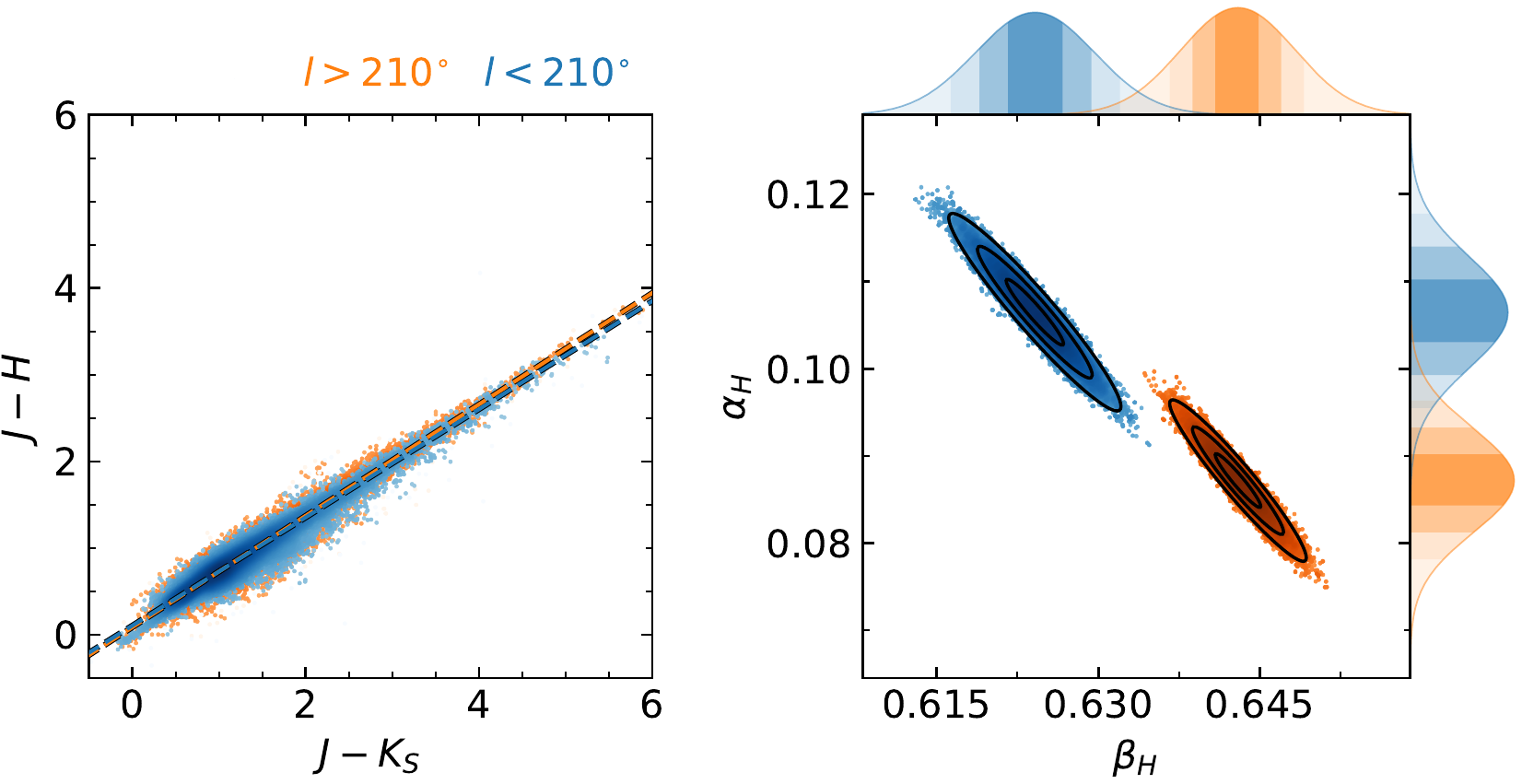}}
        \caption[]{Similar to Fig.~\ref{img:total_H}, we here show the fitting results for $H$ when splitting the survey region in east-west direction along $l=210\degr$. Our MCMC fitting method determines a significant difference in the colour excess ratios for the two sub-regions, indicating a variable extinction law across the molecular cloud.}
    \label{img:split_H}
\end{figure}

To include the above mentioned additional dispersion factor, we measure the natural width of the main sequence in the extinction-free control field for each colour. Here, we use the previously defined filtering criteria (the list above) on the control field data and add the variances accounting for the natural broadening to the corresponding squared photometric errors before running the fitting procedure. The variances of our dataset, $\sigma_{x,i}^2$ and $\sigma_{y,i}^2$ (Equ.~\ref{equ:dataset}), therefore become
\begin{align}
    \sigma_{x,i}^2 &= \sigma_{x,i,\mathrm{phot}}^2 + \sigma_{x,i,\mathrm{intrinsic}}^2 \\
    \sigma_{x,i}^2 &= \sigma_{y,i,\mathrm{phot}}^2 + \sigma_{y,i,\mathrm{intrinsic}}^2
,\end{align}
where the subscript ``phot'' refers to the measured errors of the photometry and ``intrinsic'' to the width of the intrinsic colour distribution as measured in the control field. The resulting total variances become rather large due to this additional systematic term, which, however, is well balanced by the large amount of available data points.

In addition to these modifications, we furthermore only include sources satisfying $J - K_S >$ \SI{1}{mag} in the fit. This value represents the typical red limit in this colour for unextincted late-type main sequence stars and therefore ensures that only sources which are at least minimally affected by extinction are included in the fit. Furthermore, we define the prior function (Equ. \ref{equ:prior}) with generous limits for the slope and the intercept as
\begin{equation}
\label{equ:prior:ccd}
\ln P(\alpha, \beta) =
  \begin{cases}
    0       &   -5 \leq \alpha \leq 5 ~ \land ~ -10 \leq \beta \leq 10 \\
    -\infty &   \mathrm{otherwise}
  \end{cases}
.\end{equation}
This definition allows very large ranges in both parameters and at the same time helps to speed up the convergence in the Markov chain.

\begin{table*}
    \caption{Fitting results for the colour excess ratio $\beta_{m_{\lambda}} = E_{J-m_{\mathrm{\lambda}}} / E_{J - K_S}$ and the intercept $\alpha_{m_{\lambda}}$. Due to this definition, the values for $J$ and $K_S$ are fixed. Tabulated are the average values for the cloud (Sect. \ref{sec:results:average}), results from the literature ($\beta_{m_{\lambda}, \mathrm{lit}}$) including the reference, as well as the fitting results when splitting the survey region at $l=210\degr$ into east and west parts (Sect. \ref{sec:results:spatial}). Also the relative difference in $\beta_{m_{\lambda}}$ between east and west regions ($\Delta\beta$) is given. The total extinction ratios relative to the $K_S$ band and the associated errors have been calculated using Equ. \ref{equ:a_lambda} and adopting $A_J / A_{K_S} = 2.5 \pm 0.15$ mag.}
    \label{tab:law}
    \begin{tabular*}{\linewidth}{l @{\extracolsep{\fill}} c c c c c c c c c c c c}
        \hline\hline
Band	&	$\langle \alpha_{m_{\lambda}} \rangle$	&	$\langle \sigma_{\alpha} \rangle$	&	$\langle \beta_{m_{\lambda}} \rangle$	&	$\langle \sigma_{\beta} \rangle$	&	$\beta_{m_{\lambda}, \mathrm{lit}}$	&	$\langle A_{m_{\lambda}} / A_{K_S} \rangle$	&	$\langle \sigma_{A_{m_{\lambda}} / A_{K_S}} \rangle$	&	$\beta_{m_{\lambda}, \mathrm{east}}$	&	$\sigma_{\beta, \mathrm{east}}$	&	$\beta_{m_{\lambda}, \mathrm{west}}$	&	$\sigma_{\beta, \mathrm{west}}$	&	$\Delta\beta$			\\
	&	(mag)	&	(mag)	&		&		&		&		&		&		&		&		&		&	(\%)			\\
\hline
$z$	&	-0.306	&	0.009	&	-1.371	&	0.007	&	-1.560\tablefootmark{1}	&	4.56	&	0.29	&	-1.372	&	0.009	&	-1.334	&	0.012	&	2.8	$\pm$	1.1	\\
$y$	&	-0.410	&	0.008	&	-0.925	&	0.007	&	-0.920\tablefootmark{1}	&	3.89	&	0.24	&	-0.930	&	0.009	&	-0.903	&	0.011	&	3.0	$\pm$	1.6	\\
$J$	&	-	&	-	&	0.000	&	-	&	-	&	2.50	&	0.15	&	0.000	&	-	&	0.000	&	-	&		-		\\
$H$	&	0.094	&	0.002	&	0.636	&	0.002	&	0.640\tablefootmark{2}	&	1.55	&	0.10	&	0.643	&	0.002	&	0.624	&	0.003	&	-3.0	$\pm$	0.6	\\
$K_S$	&	-	&	-	&	1.000	&	-	&	-	&	1.00	&	-	&	1.000	&	-	&	1.000	&	-	&		-		\\
$I_1$	&	-0.070	&	0.014	&	1.239	&	0.008	&	1.243\tablefootmark{3}	&	0.64	&	0.09	&	1.224	&	0.010	&	1.264	&	0.012	&	3.2	$\pm$	1.2	\\
$I_2$	&	-0.096	&	0.023	&	1.297	&	0.012	&	1.307\tablefootmark{3}	&	0.56	&	0.09	&	1.281	&	0.016	&	1.321	&	0.019	&	3.0	$\pm$	1.8	\\
$I_3$	&	-0.059	&	0.023	&	1.335	&	0.012	&	1.331\tablefootmark{3}	&	0.50	&	0.09	&	1.323	&	0.015	&	1.358	&	0.021	&	2.6	$\pm$	1.9	\\
$I_4$	&	-0.052	&	0.021	&	1.324	&	0.011	&	1.329\tablefootmark{3}	&	0.51	&	0.09	&	1.329	&	0.013	&	1.319	&	0.024	&	-0.8	$\pm$	2.1	\\
$M_S$	&	-0.172	&	0.089	&	1.364	&	0.044	&	1.370\tablefootmark{3}	&	0.45	&	0.11	&	1.376	&	0.069	&	1.342	&	0.059	&	-2.5	$\pm$	6.8	\\
\hline
    \end{tabular*}
\tablebib{
(1)~\citet{schlafly11}; (2) \citet{indebetouw05}; (3) \citet{flaherty07}.
}
\end{table*}

We note here, that in our case for accurate determinations of the colour excess ratios, we also need to allow the intercept $\alpha_{m_{\lambda}}$ to be a free parameter in the fit. This is necessary since each colour is characterized by a different intrinsic mean, leading to asymmetric offsets in the colour-colour diagrams. Furthermore, we also note that the actual dispersion along the reddening vector is also a function of extinction: only intrinsically bright background sources (early spectral types) will be visible in regions of high column-density, thus reducing this systematic dispersion for increasingly redder colours. However, instead of introducing more systematic errors by adding an additional model component, we chose to be conservative and use the same intrinsic colour dispersion for all sources in a given colour combination.

\subsection{The average extinction law}
\label{sec:results:average}

After applying the filtering steps as listed above to the data collection, we first derived an average extinction law for the entire \oriona molecular cloud by fitting all available data in each band. The results of this procedure are (among others) tabulated in Table \ref{tab:law} for each band, where the measured colour excess ratios $\beta_{m_{\lambda}}$ are also converted to total extinction ratios $A_{m_{\lambda}} / A_{K_S}$ using Equ. \ref{equ:a_lambda} and $A_J / A_{K_S} = 2.5 \pm \SI{0.15}{mag}$ adopted from \citet{indebetouw05}. With this equation it also becomes trivial to convert the measured slopes into total extinction ratios, in case a different zero-point is desired. As a representation for all bands, Fig.~\ref{img:total_H} shows the fitting results for the $H$ band. Also displayed are the posterior probability distributions of the colour excess ratio $\beta_H$ (the slope) and the intercept $\alpha_H$. In the specific of case of $H$, we find $\langle \beta_H \rangle = 0.636 \pm 0.002$, thus a statistical error well below 1\%. Upon converting this value to the total extinction ratio, we also take the errors from \citet{indebetouw05} into account. We then find $A_H/A_{K_S} = 1.55 \pm 0.1$.

Due to the introduction of this additional error component when converting the fitted slope to total extinction ratios, it is far better to directly compare the colour excess ratios $\beta_{m_{\lambda}}$ to results available in the literature. For the Pan-STARRS bands $z$ and $y$, we convert the results of \citet{schlafly11} to $\beta_{z,\mathrm{lit}} = -1.56$ and $\beta_{y,\mathrm{lit}} = -0.92$. Compared to our values of $\beta_z = -1.371$ and $\beta_y = -0.925$, only $y$ band matches, while for $z$ we find a clear deviation. This, however, can be caused by the different normalization as \citet{schlafly11} list the extinction for the other NIR bands in the UKIRT photometric system. In contrast to our data, the UKIRT $K$ band is slightly different compared to VISTA $K_S$.

In the NIR, literature results generally agree very well with our value. Most prominently, the often-cited work by \citet{indebetouw05} and \citet{wang14} both find $\beta_{H,\mathrm{lit}} = 0.64$. Other results for this specific value are from \citet{martin90}: 0.63; \citet{nishiyama06}: 0.64; \citet{stead09}: 0.65; \citet{wang13}: 0.65. All of these are in excellent agreement with our findings and reinforce our conviction that the chosen fitting model and the filtering procedure accurately describe the colour excess ratios. Also, we emphasize here that the calibration towards total extinction ratios with the results from \citet{indebetouw05} is not in conflict with these findings, because we derive the slope $\beta$ prior to the conversion.

\citet{flaherty07} already derived the total extinction ratios for \oriona in the IRAC bands with data from the \textit{Spitzer} Orion programme. In contrast to our study, they used the less sensitive 2MASS observations to fit the colour excess ratios and therefore have fewer data points for their analysis. Nevertheless, also here, our values are in excellent agreement with their result of $\beta_{I_1, \mathrm{lit}} = 1.243$, $\beta_{I_2, \mathrm{lit}} = 1.307$, $\beta_{I_3, \mathrm{lit}} = 1.331$, and $\beta_{I_4, \mathrm{lit}} = 1.329$. For $M_1$, they did not derive the extinction for \oriona, but find $\beta_{M_1, \mathrm{lit}} = 1.37$ and 1.32 for Serpens and NGC 2068/2071, respectively, which also agrees very well with our findings.

\begin{figure*}
    \centering
    \includegraphics[width=\hsize]{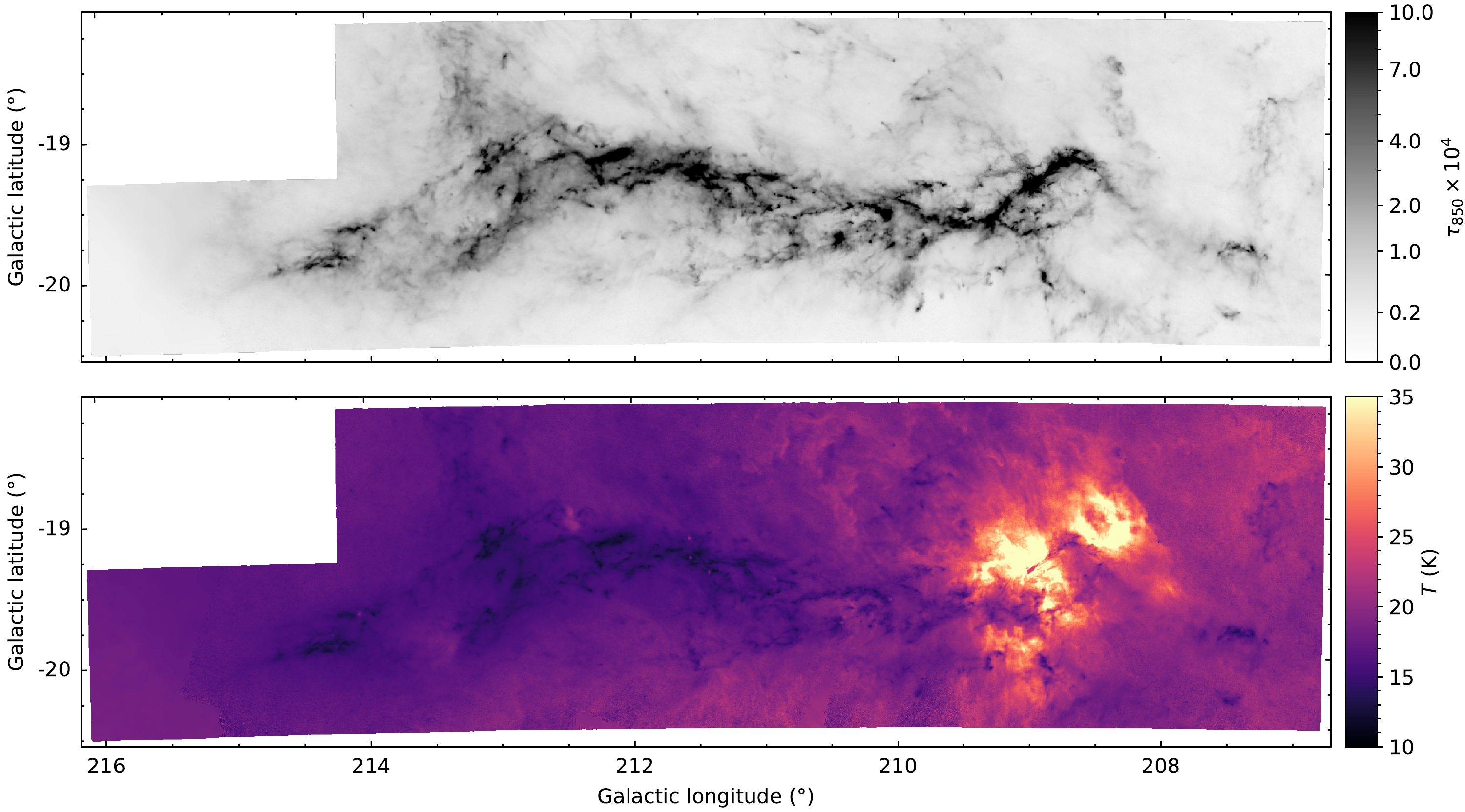}%
    \hspace{-\hsize}%
    \begin{ocg}{img:regions_a}{img:regions_a}{0}%
    \end{ocg}%
    \begin{ocg}{img:regions_b}{img:regions_b}{1}%
        \includegraphics[width=\hsize]{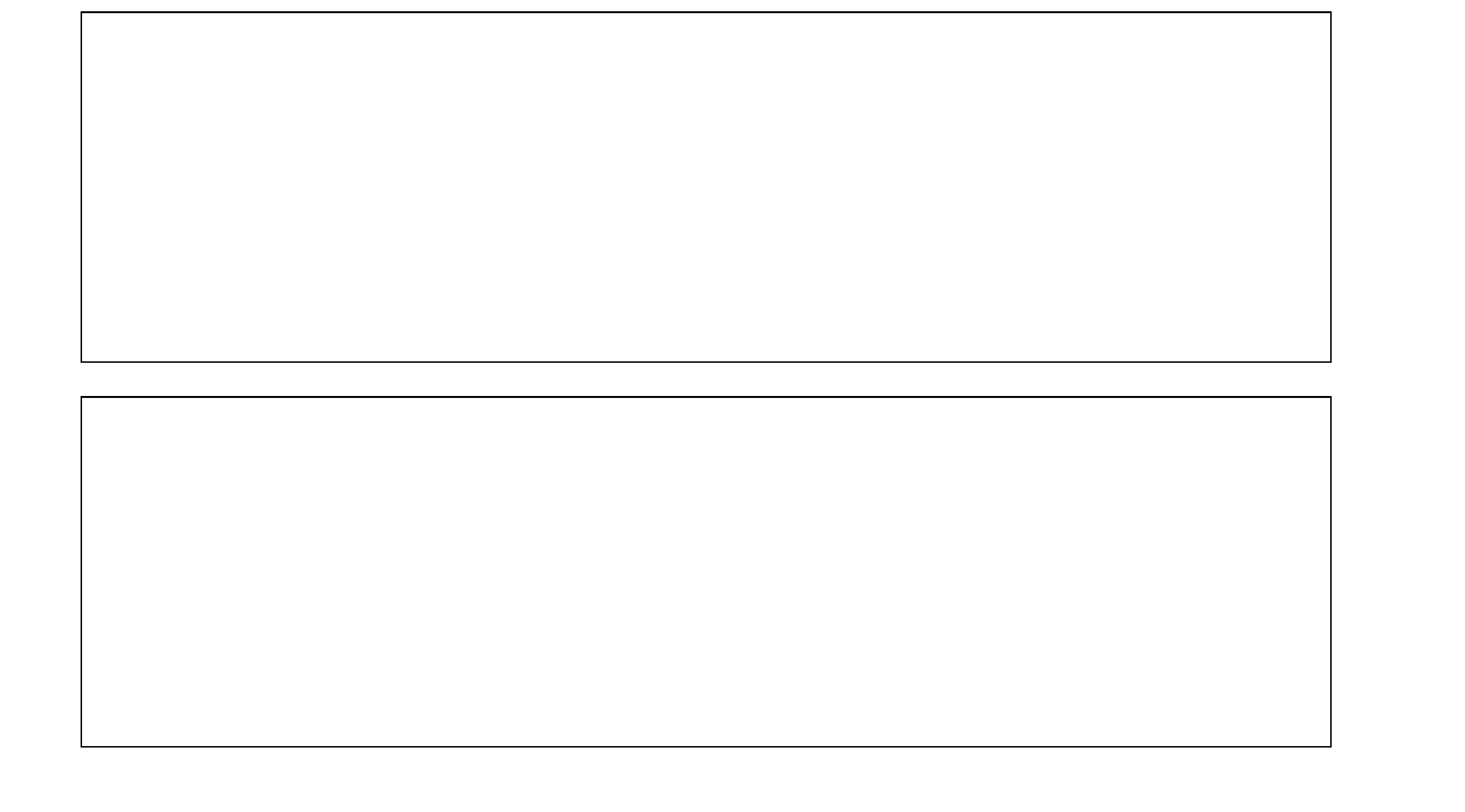}%
        \hspace{-\hsize}%
        \includegraphics[width=\hsize]{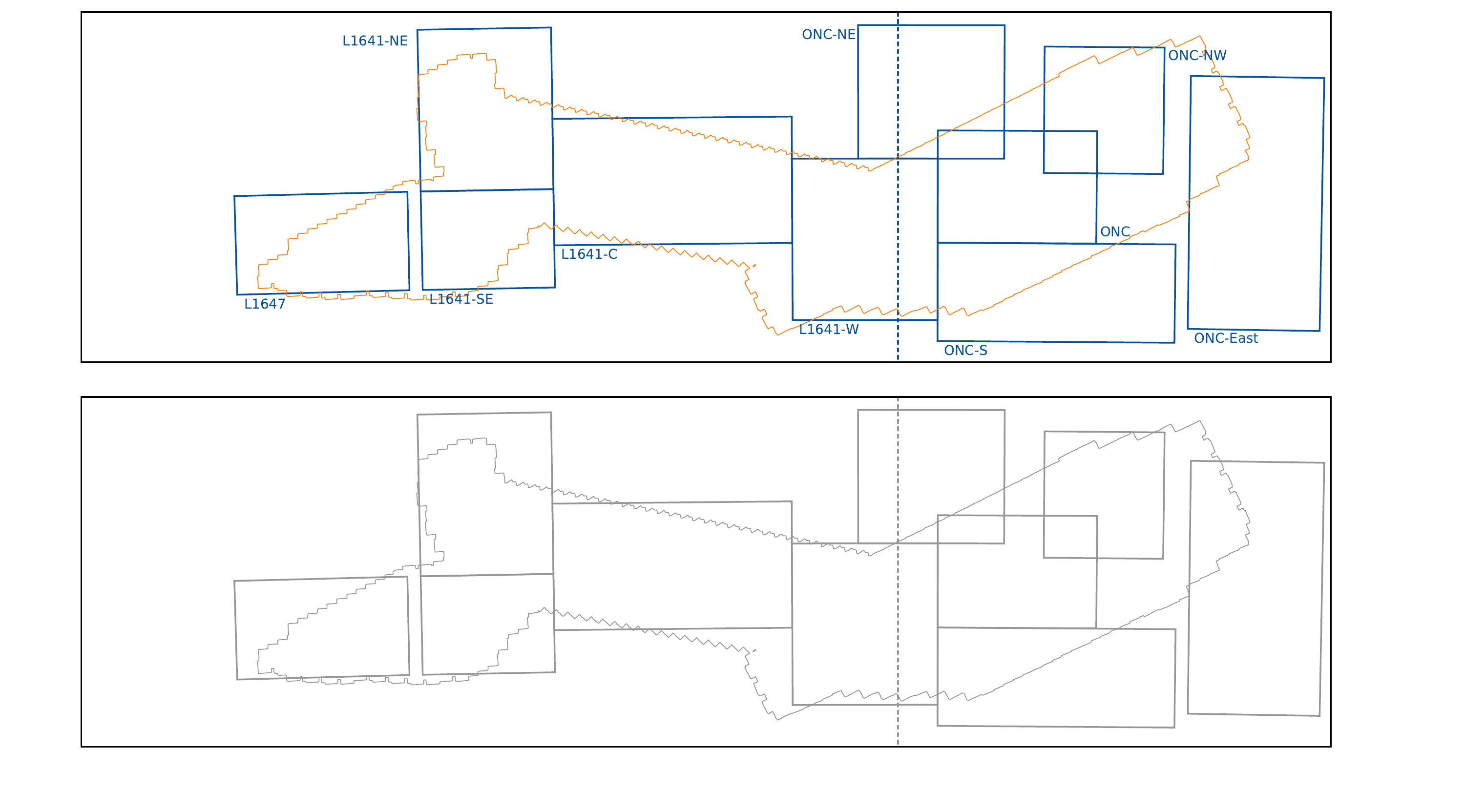}%
    \end{ocg}%
    \caption{Our sub-region layout to study variations in the colour excess ratio. The top panel shows a dust optical depth map, the bottom panel the dust temperature derived by fitting a modified black-body to \textit{Herschel} and \textit{Planck} far-infrared data \citep{lombardi14cor, lombardi14}. The analysis for the sub-regions is split into two parts: First we study variations of the colour excess ratio by splitting the cloud in east-west direction at $l=210\degr$ (dashed line). Subsequently, the entire cloud is split into multiple sub-regions, outlined here by the boxes drawn with solid lines. Regions are labelled only in the top panel and the \textit{Spitzer} survey contours for $I_1$ and $I_2$ are also drawn. \ToggleLayer{img:regions_b,img:regions_a}{\protect\cdbox{toggle labels}}}
    \label{img:extinction_law_regions}
\end{figure*}

\begin{figure*}
        \centering
        \hspace*{3.5cm}\includegraphics[width=15cm,height=10cm,keepaspectratio]{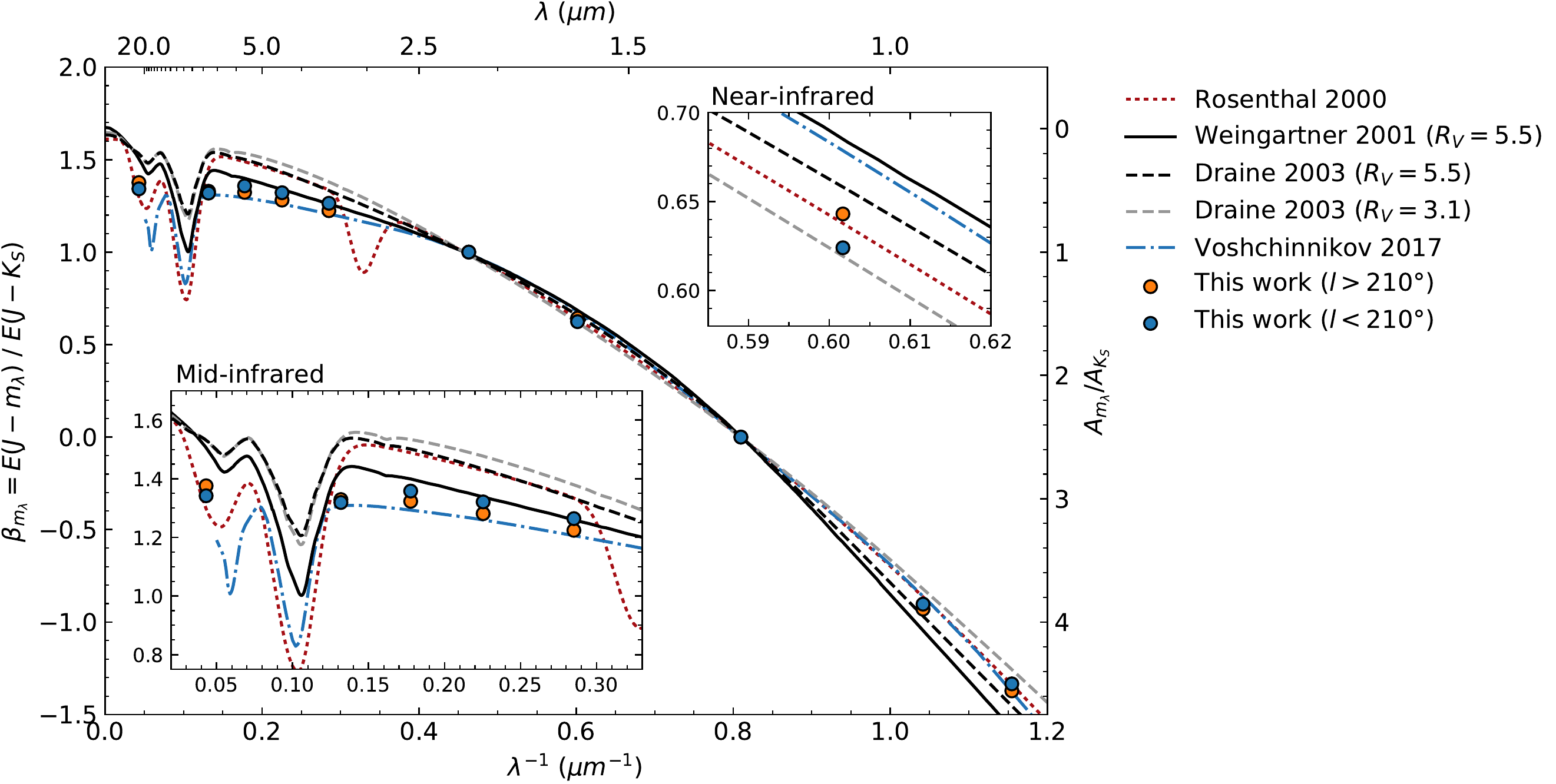}
    \caption[]{Infrared extinction curve of \oriona in terms of the colour excess ratio $\beta_{m_{\lambda}}$ as used in this manuscript. The total extinction ratio $A_{m_{\lambda}}/A_{K_S}$ on the right axis of the plot is the conversion of this ratio as given in Equ. \ref{equ:a_lambda} with values adopted from \citet{indebetouw05}. The extinction law as predicted from models by \citet[][WD01]{weingartner01}, \citet[][D03]{draine03}, and \citet[][V17]{voshchinnikov17} is displayed with various lines. The red dotted line is the fit of \citet{rosenthal00} to their data for OMC-1 near the ONC (including the \SI{{\sim}3.1}{\micro\metre} ice feature). The blue and orange circles show our measurements for the east-west split of the survey region.}
    \label{img:extinction_law}
\end{figure*}

\subsection{Spatial variations across the cloud}
\label{sec:results:spatial}

The results presented in this section are split into two parts. Firstly, we investigate how the extinction law varies across the projected surface of the molecular cloud and, in particular, between regions which are affected by the intense radiation from the massive stars in the ONC. Secondly, we test whether the extinction law is constant when fitting regions with variable gas densities.

\subsubsection{Dependence on environment}

Among all nearby star-forming regions, the \oriona molecular cloud presents a unique opportunity to study variations in the extinction law across different environments. While the western\footnote{In this manuscript, we always refer to the galactic coordinate frame when using cardinal directions.} parts of the cloud are affected by feedback from the hot, ionizing radiation of the massive stars, located in or near the ONC, to the east we find much more quiescent regions with smaller groups of YSOs and isolated star-forming events. Thanks to the supreme sensitivity of the VISTA observations, it becomes possible to systematically study changes in the colour excess ratios for individual regions. With about ten times more detected sources than previous surveys, the new $JHK_S$ data are generally a much better match in sensitivity to the \textit{Spitzer} data compared to 2MASS, enabling us to cross-validate our findings with multiple bands.

Despite sharing a visually very similar distribution, the MCMC procedure finds a statistically significant difference for the colour excess ratio, implying a variable extinction law. For $H$, we find a 3-$\sigma$ significance in the differences of the two slopes, while for $z$, $y$, $I_1$, and $I_2$ we find 1-$\sigma$ differences. The other bands ($I_3$, $I_4$, $M_1$), unfortunately, do have have the necessary sample sizes to infer a significant difference within the errors. Taking a closer look at the left-hand side panel of Fig.~\ref{img:split_H}, at first glance, it seems that there is very little difference in the distribution. Looking closer, however, it is possible to see a small difference in the distribution. Even to the eye, the $l>210\degr$ sample (orange) shows a trend towards larger $J-H$ values with increasing $J-K_S$ as the data systematically envelope the other sample at the upper edge of the distribution (e.g. at $J-K_S \approx 3$, $J-H \approx 2$ mag).

The right-hand side panel of Fig.~\ref{img:split_H} shows the posterior probability distributions of the linear fit parameters for both sub-regions. Interestingly, both results are found close to the edge or partially outside the average distribution of Fig.~\ref{img:total_H}. In other words, the average distribution assigns 0 probability to many possible parameter combinations derived for each sub-region individually, which seems to lead to inconsistent results at first. However, this can be explained in the nature of the linear fits. The MC algorithm randomly samples the parameters space ($\alpha$ and $\beta$) for all data points and does not randomly sample the data space. In contrast, the two fits in Fig.~\ref{img:split_H} refer to two distinct subsets in data space. The fact that 0 probability is assigned in Fig.~\ref{img:total_H} to both high and low values from Fig.~\ref{img:split_H} is therefore a natural consequence of fitting all data points simultaneously.

Nevertheless, even though our linear fitting setup is already very conservative regarding statistical and systematic errors, at this point one can still be very skeptical about our findings. Therefore, as verification of this result, we compared the fitted colour excess ratios for all bands to a variety of model predictions. For optimal comparison, we converted all model values to our slope convention, $\beta_{m_{\lambda}} = E_{J-m_{\mathrm{\lambda}}} / E_{J - K_S}$, by integrating the available tables over the filter transmission curves.

Figure~\ref{img:extinction_law} shows our fitted slopes as a function of wavelength on top of the models from \citet[][hereinafter WD01]{weingartner01}, \citet[][D03]{draine03}\footnote{The models from WD01 and D03 are available for download at \href{https://www.astro.princeton.edu/~draine/dust/dustmix.html}{https://www.astro.princeton.edu/~draine/dust/dustmix.html}}, and \citet[][model 25; V17]{voshchinnikov17}, as well as the data from \citet{rosenthal00}. These models represent a mixture of dust particle characteristics and size distributions. The D03 models are based on case A of \citet{weingartner01} which consists of carbonaceous and silicate grains with sizes ranging from a few \SI{}{\angstrom} to several \SI{}{\micro\metre}. D03 adjusted the size distribution for these models where for $R_V=3.1$ the grain abundance was reduced, while for $R_V=5.5$ the abundance was increased. The WD01 $R_V=5.5$ model in Fig.~\ref{img:extinction_law} refers to case B in their work case B contains a significant fraction of very large carbonaceous dust grains (1 -- \SI{10}{\micro\metre}), while case A stops at around \SI{1}{\micro\metre}. The work by V17 is based on laboratory measurements of optical properties of three-layered spheres with the specific aim to understand the widely observed flat MIR extinction law. The comparison to \citet{rosenthal00} is especially interesting as they measure an infrared extinction curve towards \object{OMC-1} which lies very close to the ONC.

One particularly interesting observation of our results refers to the difference in the colour excess ratio between the east and west regions, which we also list in Table. \ref{tab:law} ($\Delta\beta$). Here, we find that for the east-west split, all passbands up to $I_3$ show a deviation of $\sim$3\%. For the passbands $z$, $y$, $I_1$, $I_2$, and $I_3$ we find a positive difference, while for $H$ this difference is negative. More specifically, for the region including the ONC and its surroundings ($l < 210\degr$), we see larger $\beta_{m_{\lambda}}$ values for the Pan-STARRS and the \textit{Spitzer} MIR bands, while in $H$ $\beta_{m_{\lambda}}$ is smaller. In reference to Fig.~\ref{img:extinction_law}, we note that the same characteristic is also present in the dust models. Considering for the moment only the models of WD01 and D03, we can see a different behaviour of  $\beta_{m_{\lambda}}$ when comparing the $R_V=3.1$ to the $R_V=5.5$ models: While for $z$, $y$, and the MIR \textit{Spitzer} bands, $\beta_{m_{\lambda}}$ is systematically smaller for $R_V=5.5$ compared to the 3.1 models, in $H$ this trend is reversed (we note that due to our colour excess ratio definition we have fixed values of $\beta_{m_{\lambda}}$ for $J$ and $K_S$. Only for $H$ band we derive an independent estimate of the slope). Excluding the $I_4$ and $M_1$ channel with their much larger uncertainties, we see the exact same behaviour in our fitted colour excess ratios. This comparison therefore serves as a critical reinforcing argument with respect to the statistical significance of our linear fits. If our derived values were to be affected by an additional systematic noise component (e.g. biased photometry near the ONC due to extended emission), we would not expect to see such an agreement in the trends, but rather a random distribution. The slopes would also remain the same if, for instance, only the $H$ band was affected by a systematic offset in the photometry. Such an offset in one band only leads to a shift in the colour-colour diagram and therefore would not affect the slope.

\begin{figure}
        \centering
        \resizebox{\hsize}{!}{\includegraphics[]{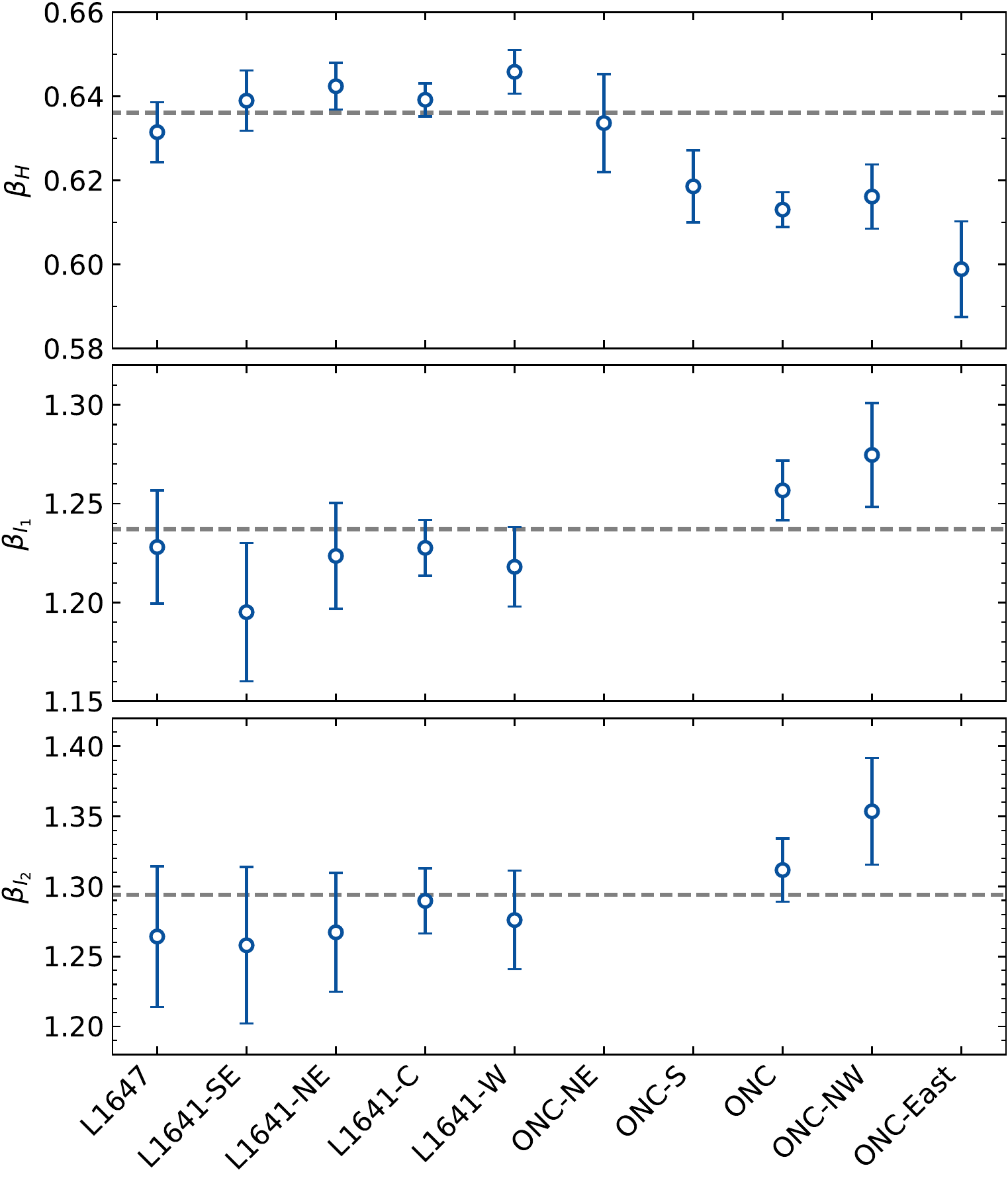}}
        \caption[]{Colour excess ratios $\beta_{m_{\lambda}}$ in $H$, $I_1$, and $I_2$ for the sub-regions of \oriona as defined in Fig.~\ref{img:extinction_law_regions}. From left to right, overall the galactic longitude of the regions decreases. The dashed horizontal lines represent the averages across the entire cloud. The slopes for \object{L1641}/\object{L1647} are systematically larger than the average in $H$ (smaller for $I_1$ and $I_2$), while for regions near the ONC, we observe the opposite behaviour.}
    \label{img:beta_subregions}
\end{figure}

To investigate the origin of these variations in more detail, we divided our dataset into even smaller sub-regions. These definitions are shown in Fig.~\ref{img:extinction_law_regions} and are adjusted so that (a) regional variations can be mapped and (b) there were still a reasonably significant amount of sources available. Due to the latter criterion, it was only possible to get reliable fits for $H$, $I_1$, and $I_2$. The other bands did not have the necessary amount of measured sources to support such an analysis. Also here we repeated the fitting procedure for each sub-region with the same setup as before. The resulting colour excess ratios for the three passbands are displayed in Fig.~\ref{img:beta_subregions}. The order of the regions in the figure is organized in such way that the galactic longitude overall decreases from left to right. In $H$ band there is a clearly visible trend towards smaller $\beta$ for decreasing longitude. Starting with the ONC-S region and continuing towards the western parts, the values of $\beta$ are systematically smaller than the average value for the cloud (the dashed horizontal line). Furthermore, the values for the \object{L1641}  regions are fairly constant and are near or slightly above the average. Within the error bars this is also true for \object{L1647}. The regions near the ONC, on the other hand, can be associated with value of $\beta_H$ smaller than the average (with the exception of ONC-NE).

\begin{figure}
        \centering
        \resizebox{\hsize}{!}{\includegraphics[]{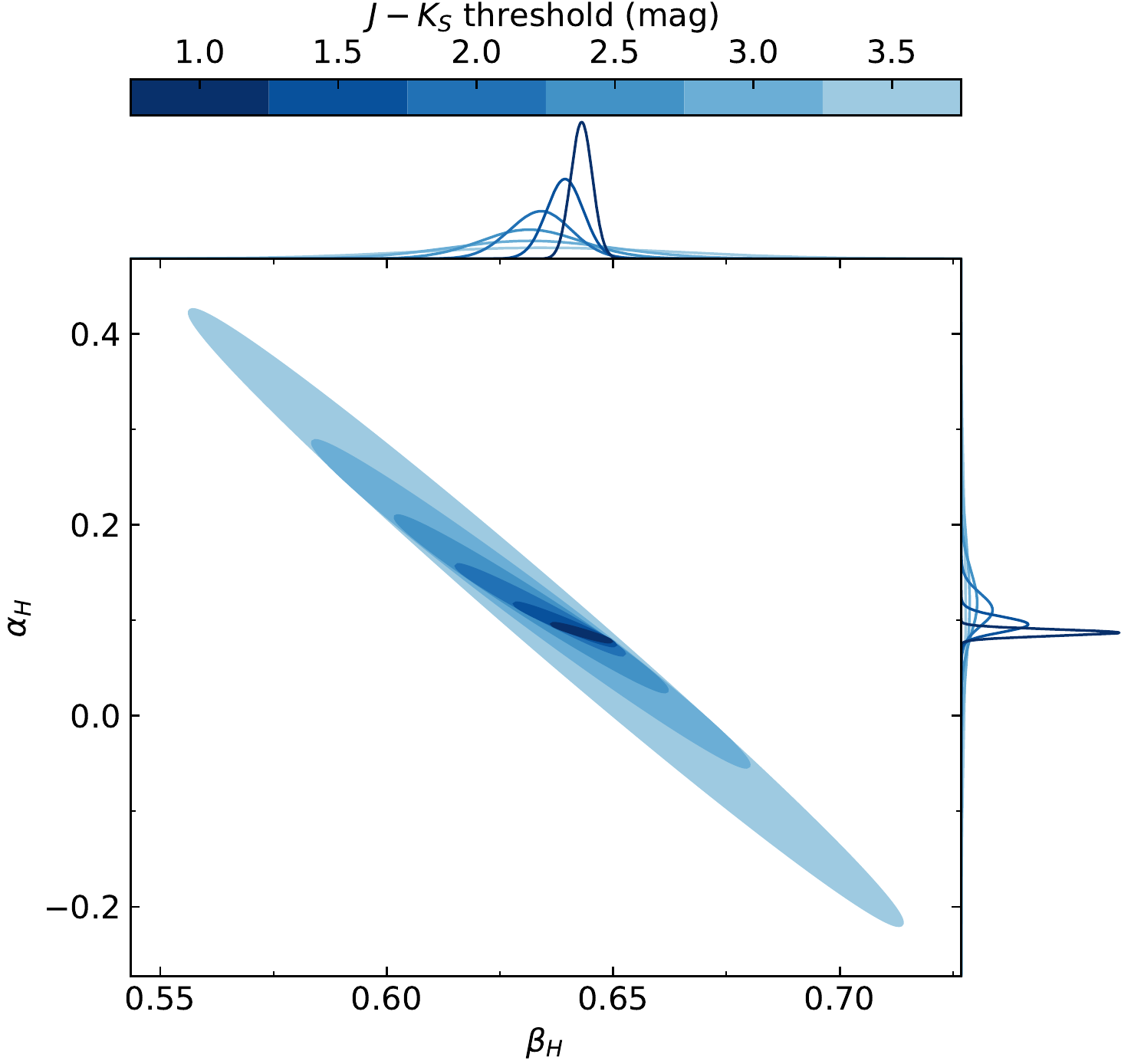}}
        \caption[]{Posterior probability distributions for the colours excess ratio $\beta_H$ and the intercept of the linear fit in the L1641/L1647 region ($l>210\degr$) when restricting the sample to various $J-K_S$ colour thresholds. For example, for the $J-K_S = \SI{1}{mag}$ threshold only source above this limit are fitted. The different thresholds are shown in increasingly dark blue colours and, similar to Fig.~\ref{img:total_H}, the ellipses represent 3-sigma covariances of the parameter distribution. The histograms represent kernel densities for each parameter of the fit.}
    \label{img:thresh_L1641_H}
\end{figure}

For $I_1$ and $I_2$, three regions (ONC-NE, ONC-S, and ONC-East) had to be omitted in this analysis, because these were only partly covered by the \textit{Spitzer} survey and did not include the necessary number of sources to produce a reliable fit. Even though the results for these two bands are not as convincing as for $H$, the general trend persists. Moreover, in agreement with model predictions, also here $\beta$ increases for the regions near the ONC, while for $H$ the colour excess ratio decreases. In light of all these findings, we finally conclude, that our results are indeed reliable and significant, which show that the infrared extinction law varies across the cloud. These variations are only of the order of 3\%, but this value is consistent across the six passbands $z$ through $I_3$.

\subsubsection{Dependence on column-density}
\label{sec:results:threshold}

Due to external environmental effects on local dust characteristics, it is also reasonable to assume that embedded regions (i.e. regions with large extinction) harbor different dust grains compared to more exposed parts of a molecular cloud. To test this dependency, we repeated the fitting procedure once again, but this time on samples with variable amounts of extinction. As extinction is directly proportional to a source's colour, we performed this test by restricting our samples with a series of colour thresholds in $J-K_S$. Because of our previous result on the variable infrared extinction law in east-west direction across the survey region, we need to take extra precautions here. Therefore, for an unbiased result, we performed this test only for the sub-region defined by $l > 210\degr$ (L1641/1647), which shows a relatively constant behaviour in the colour excess ratio. The results of the fit are visualized for $H$ in Fig.~\ref{img:thresh_L1641_H}, where the posterior probabilities for the intercept and the slope are shown. The distributions for $I_1$ and $I_2$ look almost identical with the exception of larger errors.

Across all passbands and sub-regions, and within the statistical significance of our fits, we are not able to determine a variation in the colour excess ratio as a function of $J-K_S$ colour. However, the regional variations determined above are of the order of only a few percent. The smaller sample sizes with increasing colour threshold lead to increasingly large errors in this determination for high extinction thresholds. Therefore, our sample size may not be large enough to make a final assessment on an extinction-dependent colour excess ratio in \oriona. Nevertheless, these results agree with the findings of \citet{wang14} who also concluded that the NIR extinction law does not vary with gas density.

\subsubsection{Systematic error sources}

In light of our findings of a spatially variable extinction law across the \oriona molecular cloud, it is worthwhile to discuss potential sources for systematic errors. Such errors could be attributed to, for example, source confusion, changes in the stellar populations across the field (along with variations in intrinsic colours), variable noise properties arising from differently calibrated data, or unaccounted foreground populations.

Source confusion mostly affects the region near the ONC and possibly other groups of stars with above-average spatial densities (i.e. groups of young stars). To minimize this impact, we removed all known YSOs prior to the linear fitting (see Sect. \ref{sec:xmatch}). Additionally, we verified our results by excluding sources in the ONC region (\SI{20}{\arcmin} around the given centre in the SIMBAD database). Doing so changes the derived $\beta_{m_{\lambda}}$ values only by about 1\%, but the conclusions derived from Fig. \ref{img:extinction_law} and Tab. \ref{tab:law} are unaltered. Moreover, observations of a variable colour-excess-ratio also persists in the ONC-East region where source confusion is not expected to be of any significance.

Another possible systematic error source could originate from changing stellar populations across the field. We do not expect this to be significant given the fact, that the \oriona molecular cloud is located towards the galactic anti-centre and \SI{20}{\degree} below the galactic plane. Even our control field data, located at the same galactic latitude, but \SI{20}{\degree} further east, does not show a significant change in stellar colours, when compared to the \oriona field. An observable signature of changing stellar populations, for instance, would be a shift in intrinsic colours. We tested this hypothesis by deriving intrinsic colours in bins of \SI{2}{\degree} in galactic longitude with \pnicer for the colours $J-H$ and $H-K_S$. We find a standard deviation of only \SI{2}{mmag} (0.5\%) and \SI{3}{mmag} (2\%) for the mean intrinsic colour in these two passbands, respectively. However, we have to add here that the derived errors of individual intrinsic source colours are much larger (\SI{>0.1}{mag}).

In \citetalias{meingast16}, we also highlighted other potential stellar groups in the foreground of the cloud. Since they lie in the foreground, they do not trace extinction through the molecular cloud and therefore should be removed prior to determining colour-excess ratios. In this work, however, we only removed the sources associated with \object{NGC 1980} from our data, because the determination of the other groups was based on an entirely statistical analysis. Thus, we do not have a source list which can be removed from our sample. In any case, any foreground objects show negligible extinction and are therefore irrelevant for the fitting procedure, where we generally rely on a large spread in colour-space (see e.g. Fig. \ref{img:total_H}).

We also checked whether the data we used show a gradient along galactic longitude in terms of observing date, which could hint at differences in their calibration. Also here, no significant discernible pattern was visible with respect to our findings. To test for further systematic errors and to validate the robustness of our results we investiagted whether the results are invariant with respect to magnitude cuts. To this end, we created additional subsets, where we restricted the selected sources to increasingly rigorous magnitude limits (e.g. $z < \SI{17}{mag}$, $y < \SI{17}{mag}$, $H < \SI{16}{mag}$, $I_1 < \SI{12}{mag}$). The changes in individual $\beta_{m_\lambda}$ values are found consistently below 1\% when when applying these magnitude cuts and consequently also the trends visible in Fig. \ref{img:extinction_law} persist.

\subsection{The shape of the \oriona infrared extinction curve}
\label{sec:results:law:discussion}

Studying Fig.~\ref{img:extinction_law} in more detail reveals several interesting findings with respect to the general trend in the infrared extinction. Overall, we find that our fitted colour-excess ratios are well traced by the various models. In particular, we see a flat MIR extinction curve, very similar to the ones that have been observed towards the galactic centre \citep[e.g][and references therein]{wang15}. This contrasts to the findings of \citet{rosenthal00} for the OMC-1 region. Compared to our measurements of a relatively flat MIR extinction law in the IRAC channels, their data suggests a continued decline in the extinction towards the \SI{9.7}{\micro\metre} feature. The same trend is seen in both D03 models. Our interpretation of this discrepancy is, that the work by \citet{rosenthal00} traces only very local conditions within OMC-1, which are not representative for the extinction law on larger scales in this region. However, they also state, that their derived extinction minimum at \SI{6.5}{\micro\metre} is not very well constrained. On the other hand, the WD01 $R_V=5.5$ (case B) model along with the new results from V17 are a much better match to our results in the MIR.

Agreements between a measured flat MIR extinction and the WD01 $R_V=5.5$ case B model are often interpreted as an indicator of the presence of very large grain sizes (up to \SI{10}{\mu\metre}). The comparison to the model by V17, however, shows, that this must not necessarily be the case. The V17 size distribution is based on work by \citet{hirashita14}, who investigate dust grain growth over several hundred Myrs. Their initial distribution is taken from case A of WD01, thus only featuring smaller grain sizes. Even after a few Myrs, dust particle sizes do not exceed \SI{1}{\micro\metre}, yet the new models of V17 are able to explain the flat MIR extinction curve.

In $H$, the situation is reversed. Here, the \citet{rosenthal00} and D03 $R_V=3.1$ models fit much better to our colour excess ratios and the models that fit best to our MIR findings (V17 and WD01) are by far the worst matches for this passband. Even worse, for both $M_1$ and the Pan-STARRS bands $z$ and $y$, yet another set of models seems to reproduce our findings better. It therefore seems, that these dust models are not able to fully reproduce the infrared extinction law from $\sim1$ to \SI{{\sim}25}{\micro\metre}, but are tuned to fit a narrower range. In conclusion for the overall infrared extinction law in \oriona, we argue that grain sizes are not the only necessary factor to explain variations, but more sophisticated models are required.

\begin{figure}
        \centering
        \resizebox{\hsize}{!}{\includegraphics[]{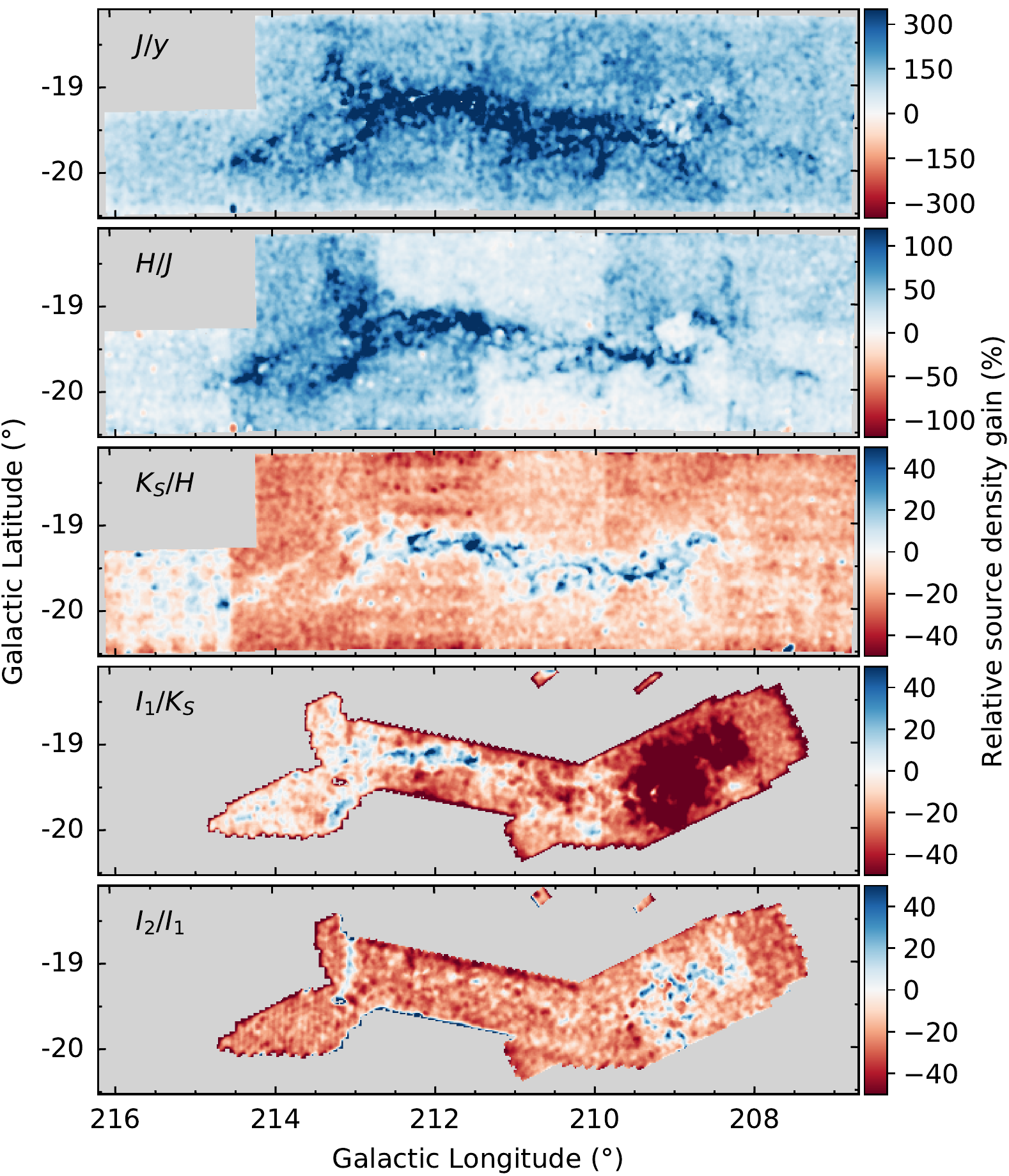}}
        \caption[]{Comparison of source densities for consecutive passbands. The individual panels show the ratio of source densities for two particular passbands (e.g. $J$ and $y$ in the topmost panel) evaluated with a \SI{1}{\arcmin} Epanechnikov kernel. The colour scale indicates the gain (or loss) of the redder band over the bluer. Clearly, the $JHK_S$ VISTA observations show overall the best sampling of sources in the region and even redder bands include more extincted sources in the cloud. Due to the small difference in the extinction between $I_1$ and $I_2$, the source density gain is negligible for $I_2$. The apparent large-scale pattern for the second and third panel from the top is caused by the variable completeness of the VISTA observations.}
    \label{img:source_gain_kde}
\end{figure}

Our analysis also revealed a variable extinction law in east-west direction across the survey region: towards the east of the ONC we find a systematically different extinction law as in regions near the cluster. Because of the above mentioned caveats when comparing our colour excess ratios to dust models, it is almost impossible to arrive at a conclusion with respect to a variation in specific dust characteristics. As disappointing as it is, for the moment we can only conclude that the grain population in general changes across the cloud. Whether this is due to grain sizes, different particle structures, or abundances, we can not say.

We can however, speculate on the origin of these variations. The bottom panel of Fig.~\ref{img:extinction_law_regions} highlights the dust temperature derived with modified black body fits to \textit{Herschel} and \textit{Planck} dust emission measurements. In this figure it is clearly visible that the \object{L1641}/\object{L1647} regions have a significantly lower dust temperature (10 -- \SI{15}{K}), while the other parts are apparently heated by the hot stars in the cluster. Even in the ONC-East region, which is spatially clearly separated from the cluster region, it seems that cold dust is surrounded by a layer of much warmer material. In contrast, the L1641W region, which is (in this projected view) closer to the cluster, does not have such an apparent layer of warm dust. It therefore seems, as if L1641 sits in the shadow of the cloud and is shielded from the intense radiation originating in the massive cluster stars. Therefore, the conclusion here is, that the radiative feedback from the hot stars significantly impacts the dust grain population.

Considering that we found variations in the colour excess ratio of the order of only 3\%, we furthermore argue that research into variations of the infrared extinction law (in particular for the NIR) should be carried out in environments where systematic errors, or external effects can be controlled to at least some extent. For instance, when determining the extinction law for various sightlines near the galactic centre or in the plane, it becomes very difficult to control the influence of physically separate regions along the line of sight, potentially diluting signals of the order of only a few percent. Thus, such studies should ideally be carried out in isolated environments, such as the \oriona molecular cloud.

\begin{figure*}
        \centering
        \resizebox{\hsize}{!}{\includegraphics[]{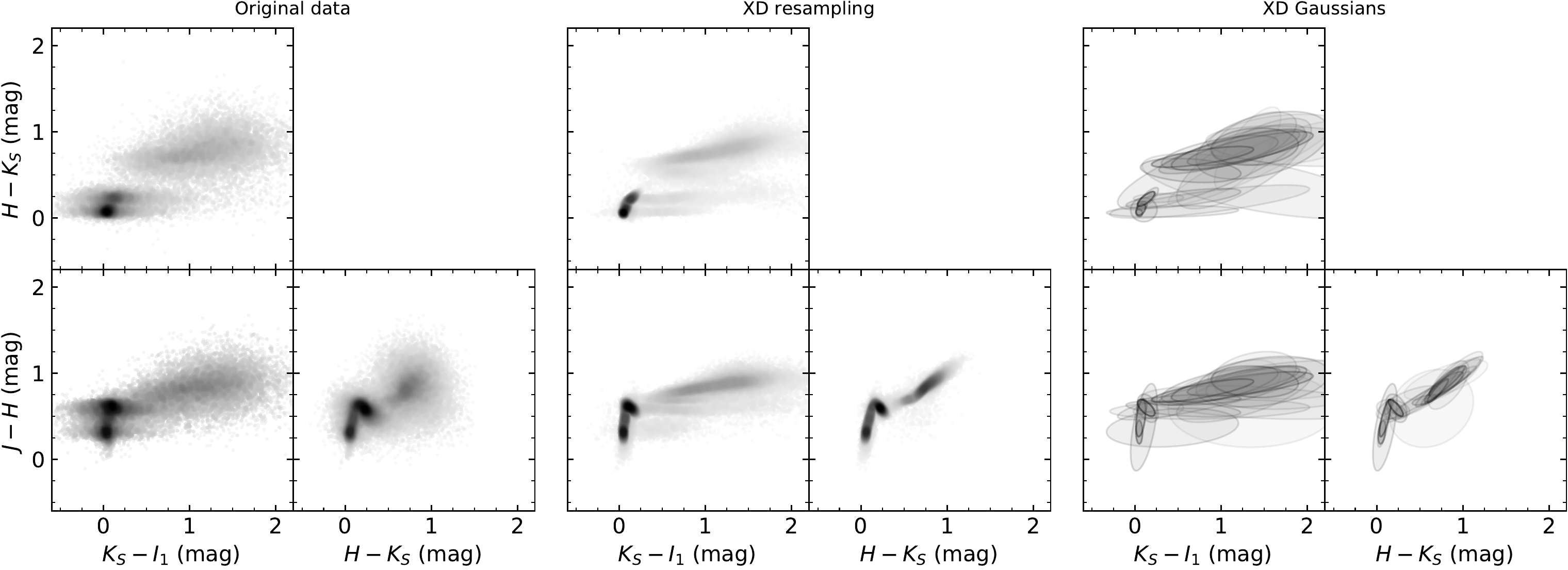}}
        \caption[]{Extreme deconvolution (XD) resampling process of the three-dimensional control field data space ($J-H$, $H-K_S$, $K_S-I_1$). The leftmost set of panels shows the original data distribution as projected views. The three panels in the centre show the same set of combinations for the resampled data with $10^5$ randomly drawn sources. The rightmost panels show the locations of the individual three-dimensional Gaussians. For these, the opacity of the ellipses is an indicator of their respective weight in the model.}
    \label{img:XD}
\end{figure*}

\begin{figure*}
        \centering
        \resizebox{\hsize}{!}{\includegraphics[]{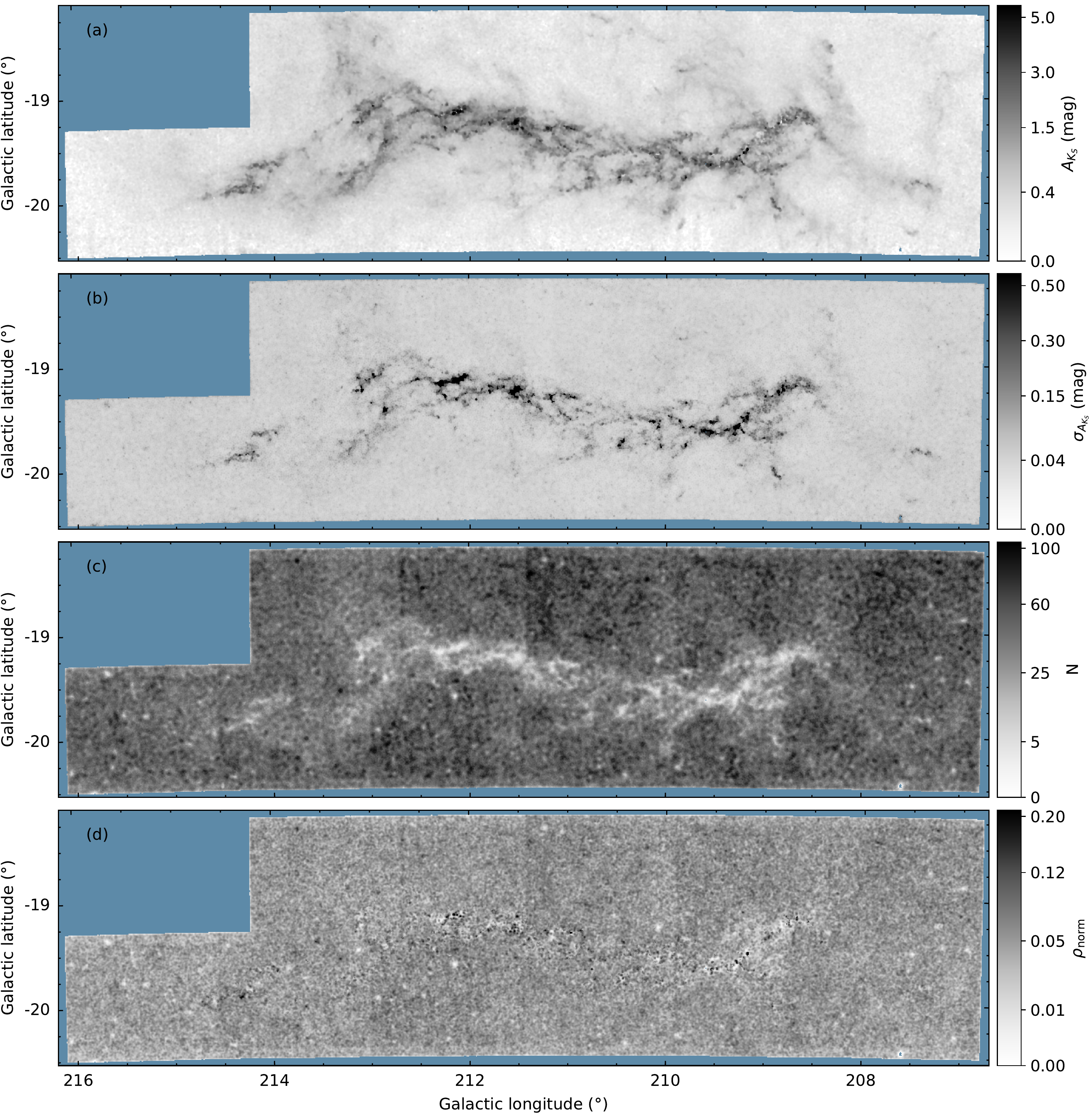}}
        \caption[]{\oriona extinction map constructed with VISTA and \textit{Spitzer} data. At the top, panel (a) shows the extinction map in $K_S$ magnitudes with a resolution of \SI{1}{\arcmin}. Panel (b) displays the associated 1-$\sigma$ errors and (c) the number of sources used for each pixel to calculate the average extinction. The bottom panel (d) shows the source kernel density when adjusted with the \nicest technique, which applies a correction factor to account for a bias in the measured extinction due to cloud substructure. Due to this correction, the cloud structure mostly disappears when compared to panel (c). The kernel density values shown in the bottom panel are normalized to the maximum density.}
    \label{img:extinction_map}
\end{figure*}

\section{\oriona extinction map}
\label{sec:map}

Following the determination and discussion on the extinction law in the previous sections, this section concentrates on creating a new extinction map for \oriona. We start by describing several required preparatory steps, followed by a detailed description of the extinction map. We close this sections by comparing our new extinction map to dust emission measurements made with \textit{Herschel}.

\subsection{Preparations}

From our analysis of the variability of the colour excess ratio it is not entirely clear which regions exactly are characterized by a given infrared extinction law. We were able to determine a trend in east-west direction which is most likely caused by the feedback from the massive cluster stars. However, drawing an exact border is not possible. For this reason we construct an extinction map for the entire survey region with the average extinction law determined in Sect. \ref{sec:results:average} (see Table \ref{tab:law}).

To calculate the individual colour excesses for each source in our catalogue, we use the \pnicer method (see Sect. \ref{sec:methods:pntsrc} for details) which is able to combine photometric measurements in arbitrary numbers of passbands. Usually, including more passbands helps to decrease the statistical noise in the extinction calculations. However, the more bands included, the more susceptible the method becomes to systematic errors. For example, in our case the Pan-STARRS photometry is often affected by extended emission, leading to systematic offsets in the measured magnitudes which are not accounted for by the statistical error from the photometry. Furthermore, in the case of our combined photometric master catalogue, it may not even be necessary to include some bands due to very different sensitivity limits and the number of available sources in general.

To find an optimal compromise, we investigated the source densities across the survey area for each band. The results for a subset of these is shown in Fig.~\ref{img:source_gain_kde}, where the gain in number density is plotted for consecutive passbands. Easily visible is the tremendous source density gain between the Pan-STARRS $y$ and the VISTA $H$ band. Not only do the NIR VISTA bands suffer much less from extinction (Fig.~\ref{img:passbands}), but also the survey in general is tuned to observe much fainter objects. Another significant gain in the entire region is achieved when using $H$ over $J$ band. For $K_S$ over $H$ and $I_1$ over $K_S$ we find significantly more sources in the highly extincted regions. From the first to the second IRAC channel there is almost no gain due to the very small difference in extinction in these two bands (see Fig.~\ref{img:extinction_law}).

In light of these findings, we chose to include only the NIR bands $J$, $H$, and $K_S$, as well as $I_1$ in the MIR. We specifically chose to not include $I_2$ due to the small length of the extinction vector in $I_1 - I_2$, which can lead to large systematic errors when determining the colour excess with just these two bands. Moreover, $I_2$ does not add sources to the highly extincted regions compared to $I_1$. The Pan-STARRS bands are omitted, because they include significantly fewer sources than the VISTA observations and do not sample the high column-densities in the first place.

\pnicer relies on intrinsic colour determinations in a control field. As outlined in Sect. \ref{sec:data}, we do not have access to suitable \textit{Spitzer} control fields data and therefore use transformed \textit{WISE} photometry (equations \ref{equ:sw1} and \ref{equ:sw2}). However, as the \textit{WISE} data are associated with larger uncertainties, the intrinsic colour distribution is expected to be significantly broader compared to \textit{Spitzer} data. This is readily visible in the comparison of the \oriona data to the control field in Fig.~\ref{img:combinations_scatter}. Furthermore, it would also be beneficial to reduce the scatter in the NIR data due to photometric errors for a better description (i.e. narrower sequences) of the intrinsic colours.

For this purpose we fit Gaussian Mixture Models to the density distributions in the three-dimensional colour space ($J - H$, $H - K_S$, $K_S - I_1$) incorporating the measurement errors. In Bayesian statistics in astronomy, this process is commonly referred to as extreme deconvolution \citep{bovy11}. We use the astroML\footnote{Source code and description available at \href{http://www.astroml.org}{http://www.astroml.org} and \href{https://github.com/astroML/astroML}{https://github.com/astroML/astroML}.} Python implementation \citep[][]{astroML, ivezic14} of this method and apply it to the control field data in all three dimensions simultaneously. The number of model components (Gaussians) is chosen by minimizing the Bayesian Information Criterion \citep[BIC;][]{schwarz78}
\begin{equation}
    \mathrm{BIC} = -2 \ln L + k \ln N
,\end{equation}
where $k$ is the number of model parameters, $N$ the number of sources, and $L$ the likelihood of the data under the given model. This is necessary because, one common caveat of such methods is that in principle it is possible to artificially increase the likelihood of models by increasing the number of components. The BIC definition attempts to correct for this fact by introducing a penalty term which includes this number ($k$). In our case the BIC showed a very flat distribution starting with about 5 -- 10 Gaussians. However, by visually inspecting the fitting results we found that some features of the distribution were only reproduced by relatively high numbers of Gaussians (20 -- 30). The fitting results of our final choice of 25 Gaussians is shown in Fig.~\ref{img:XD}. The cloned distribution of the XD resampling provides a very clean sample of the colour space. Moreover, the resampling process of the data space successfully produced a distribution which reduces the influence of photometric errors in the control field. We used this resampled distribution of the colour space to draw $10^6$ randomly selected sources to construct a model control field. Using \pnicer we then calculated line-of-sight extinctions for all sources which had independent measurements in at least two bands.

\subsection{Extinction map}

Before creating an extinction map one further complication was to select the correct density correction scaling factor $\alpha_{\textsc{Nicest}}$ (Sect. \ref{sec:methods:mapping}). This parameter represents the slope of the luminosity function and depends on the used bands. For NIR $JHK_S$ data this value typically is 1/3 \citep{lombardi09}. Here, however, we also include the \textit{Spitzer} $I_1$ channel which complicates the situation. Instead of trying to empirically derive the correct factor, we created a series of preliminary extinction maps with variable correction factors ($\alpha_{\nicest}$ = [0, 1] in steps of $10^{-2}$). The goal of this exercise was to calculate the extinction-corrected source density map and minimize the dispersion therein. The minimum value was found at $\alpha_{\textsc{Nicest}} = 0.25$.

One major disadvantage of extinction mapping with discrete line-of-sight samples is the method's dependency on the number of available background sources. The number density of the sources in highly extincted regions ultimately restricts the resolution of the map. Here, the case of the \oriona molecular cloud is one of the more difficult regions with respect to extinction mapping. This is because the cloud is located towards the galactic anti-centre and also well below the galactic plane, thus offering intrinsically very few background sources. For essentially all other nearby molecular clouds (e.g. the \object{Ophiuchus molecular cloud} or the \object{Pipe Nebula}) many more background sources are available, since they are projected against the galactic bulge. For these regions it is rather common to achieve much better extinction map resolutions compared to \textit{Herschel} dust emission maps. To determine this limit for the given data, we created a series of extinction maps with a variety of resolutions following the method described in Sect. \ref{sec:methods:mapping}. We found that the limit in resolution lies at about 1\arcmin. With this value only very few pixels could not be sampled in the high density regions of the cloud due to a lack of background sources. Well-sampled pixels are only achieved at lower resolutions of about 1.5\arcmin -- 2\arcmin. We therefore provide a series of extinction maps calculated at different resolutions which will be made available to the community via CDS.

Figure \ref{img:extinction_map} shows the final \oriona extinction map at \SI{1}{\arcmin} resolution in units of $K_S$ band extinction. In particular, we used a Gaussian kernel with $\mathrm{FWHM} = \SI{1}{\arcmin}$ which was truncated at 5 standard deviations and a pixel size of \SI{0.5}{\arcmin} for sufficient sampling. The figure also includes further statistics for each pixel: the extinction error (1-$\sigma$), the number of sources (N), and the corrected sources densities resulting from \nicest ($\rho$). The extinction map offers a large dynamic range at much better resolutions than the previously available map of \citet{lombardi11} and traces the dense gas structure of the cloud very well. The map tracing the number of the used sources for each pixel (N) clearly outlines the regions of high extinction. In contrast, the source kernel density map displayed in panel (d) does not highlight the molecular cloud's structure. This is because the kernel density in this panel incorporates the \nicest correction. In addition, the fact that the cloud is barely visible in this map is a direct indicator for the successful application of the \nicest algorithm. Only very localized prominent features are visible which are caused by parts in the VISTA image and the source catalogue where the correction factor is not applicable. This can be either due to extremely bright stars with halos on the images ($\rho_{\mathrm{norm}} \sim 0$), or due to stellar clusters in cloud ($\rho_{\mathrm{norm}} \gtrsim 0.2$). For such regions a density correction based on the expected number of sources not valid and leads to the residual pattern across the density map.

\begin{figure}
        \centering
        \resizebox{\hsize}{!}{\includegraphics[]{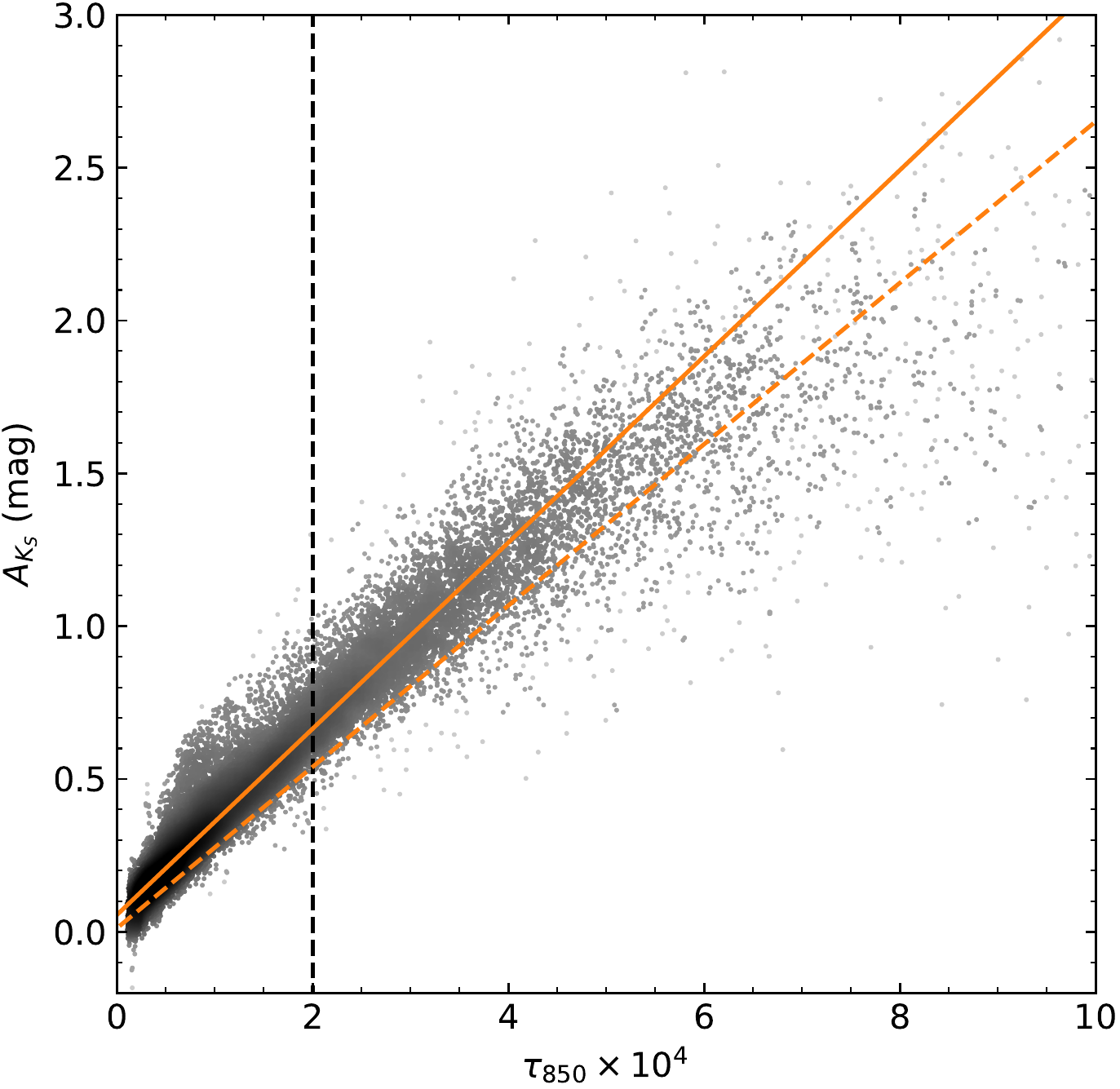}}
        \caption[]{Relation between \textit{Herschel}-\textit{Planck} dust optical depth at \SI{850}{\micro\metre} ($\tau_{850}$) and dust extinction in $K_S$. The orange solid line is our linear fit with the slope $\gamma = \SI{3042}{mag}$. The dashed orange line is the previous fit of \citet[][$\gamma = \SI{2640}{mag}$]{lombardi14} who used an extinction map based on 2MASS data. They also limited their fit to $\tau_{850} < 2\cdot10^4$, as indicated by the vertical dashed line, while we use the entire range displayed in this figure. We find a $\sim15\%$ larger slope in the relation due to much better sampling of the cloud's substructure.}
    \label{img:herschel_calibration}
\end{figure}

\begin{figure*}
        \centering
        \resizebox{\hsize}{!}{\includegraphics[]{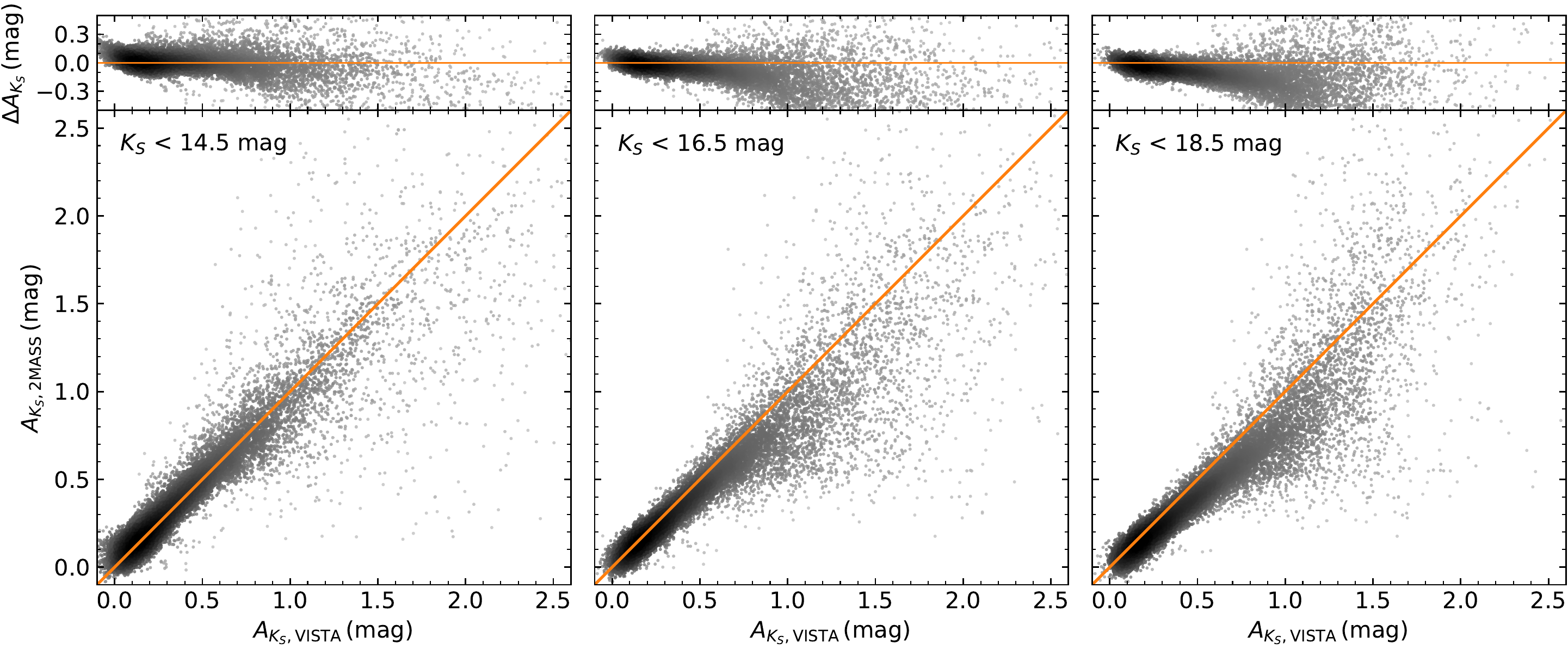}}
        \caption[]{Comparison of the pixel values for extinction maps created with 2MASS and VISTA data (sensitivity limits: \SI{{\sim}14.5}{mag} and \SI{{\sim}19}{mag}, respectively). Each of the three panel groups shows a different magnitude limit. The leftmost plot shows a comparison for extinction maps where all source brighter than \SI{14.5}{mag} in $K_S$ are used. The panel in the centre and to the right display the same comparison, but with sources brighter than 16.5 and \SI{18.5}{mag} in $K_S$, respectively. While the bottom figures directly compare the two measures, we plot their difference ($A_{K_S, \mathrm{2MASS}} - A_{K_S, \mathrm{VISTA}}$) at the top of each group. The orange lines represent 1:1 correlations between the parameters.}
    \label{img:substructure_maglim}
\end{figure*}

\subsection{\textit{Herschel-Planck} re-calibration}
\label{sec:results:herschel}

The final maximum resolution of our extinction map of \SI{1}{\arcmin} is still well above the resolution of the dust emission measurements by \textit{Herschel} ($\sim$36\arcsec). Thus, for the case of \oriona, it is practical to use the higher-resolution \textit{Herschel} dust thermal emission measurements and convert them to column-density estimates. This procedure has been performed a number of times in the recent past for various cloud complexes \citep[e.g.][]{stutz10}, but here we will use the method introduced by \citet{lombardi14}. To convert the far-infrared dust emission to column-density estimates, the authors used \textit{Herschel} and \textit{Planck} data to estimate the dust optical depth at \SI{850}{\micro\metre} ($\tau_{850}$) in the cloud by fitting modified black bodies to each pixel in the various channel maps. In a subsequent step, the resulting map of dust optical depth was converted to extinction (and therefore column-density) by cross-calibration with an extinction map. Here, the assumption was made that the dust optical depth and extinction share a linear correlation described by
\begin{equation}
    A_K = \gamma \tau_{850} + \delta
.\end{equation}
For a reliable fit, an upper limit in the parameters is necessary, because a linear correlation is expected to be only valid up to the point where the extinction map is unbiased. With typical 2MASS extinction maps this limit can be set to $\tau_{850} = 2\cdot10^4$ \citep{lombardi14, zari16}. With our new extinction map, which features a much larger dynamic range, this limit is expected to be pushed to higher values.

For fitting the relation we also use the MCMC-based method outlined in Sect. \ref{sec:methods:fitting}. In this case our prior function (Equ. \ref{equ:prior}) takes the form
\begin{equation}
\ln P(\delta, \gamma) =
  \begin{cases}
    0       &   -1 \leq \delta \leq 1 ~ \land ~ 0 \leq \gamma \leq 10^4 \\
    -\infty &   \mathrm{otherwise}
  \end{cases}
,\end{equation}
where now $\delta$ is the intercept, and $\gamma$ the slope. In this case, however, we are not able to estimate additional systematic uncertainties in the correlation and rely only on the statistical errors derived for the extinction map and the dust optical depth. For this reason, the error in our fit is severely underestimated.

In Fig.~\ref{img:herschel_calibration} we plot the relation between extinction in the $K_S$ band against the dust optical depth at \SI{850}{\micro\meter} along with our fitting result. Clearly, the linearity of the relation seems to persist beyond the previously adopted value and we fit the entire range up to $\tau_{850} = 10^5$.  We find a slope of $\gamma = \SI{3042}{mag}$, which is about 15\% steeper when compared to the previous fit of \citet{lombardi14}, who derived $\gamma = \SI{2640}{mag}$. In light of additional unknown systematic errors, we round this value to $\gamma = \SI{3050}{mag}$.

We now continue to investigate and discuss the origin of the discrepancy in the correlation of dust optical depth and extinction between the results of \citet{lombardi14} and our fit. A detailed comparison of the extinction maps from 2MASS \citep[][]{lombardi11} and our new map reveals the reason. Figure \ref{img:substructure_maglim} shows a comparison of the pixel values for \SI{3}{\arcmin} extinction maps (the resolution of the 2MASS map) created from 2MASS and VISTA data, respectively. This comparison is shown for three different magnitude limits, in the sense of only including sources brighter than the given limit to create the map. For the typical 2MASS sensitivity limit of \SI{14.5}{mag} in $K_S$, the map values share a 1:1 correlation. However, when including more sources in the VISTA map (2MASS tops out at \SI{{\sim}14.5}{mag}), a clear trend to higher extinctions in each pixel becomes visible. The most likely reason for this is that individual pixels have significant cloud substructure which only gets sampled when enough background sources are available. In the case of the typical 2MASS magnitude limit, only relatively few background sources are available per pixel. Thus, structure with a size smaller than the resolution of the map is partly not being sampled. In contrast, the VISTA data offer an order of magnitude more background sources which significantly increase the sampled area.

The maps, for which we display the pixel values in Fig.~\ref{img:substructure_maglim} have been created using the \nicest substructure correction (Sect. \ref{sec:methods:mapping}). This correction is in principal designed to minimize such an effect, however, a certain bias will still remain if structures are not sampled by background sources in the first place. It therefore seems likely, that the 2MASS observations were apparently not sensitive enough to provide a full sampling of the cloud substructure. This trend continues for even higher magnitude limits but changes become less significant when approaching the sensitivity limit of the VISTA data ($K_S \sim \SI{19}{mag}$). Still, we can not say with absolute confidence that the new data sample the substructure up to a point where no more improvements could be found by including even more background sources.

Other values for $\gamma$ between extinction and dust optical depth in the literature are published for \orionb \citep[\SI{\sim 3460}{mag};][]{lombardi14} and for the Perseus cloud complex \citep[\SI{\sim 3900}{mag};][]{zari16}. Also here the authors used 2MASS extinction maps, and both clouds share a similar unfavourable position compared to \oriona with respect to the number of available background sources. \orionb lies at about the same distance as \oriona but is closer to the galactic plane. The Perseus cloud, on the other hand lies significantly closer (\SI{240}{pc}), which reduces the problem with structure not being sampled by background sources. Nevertheless, for these clouds it seems reasonable to assume that they are also affected by this problem to a certain extent. Furthermore, we also can not exclude the possibility that the different calibrations of the \textit{Herschel} dust optical depth measurements are also affecting this relation.

The value of $\gamma$, in principle, can be correlated with physical conditions of the dust since it is proportional to the ratio of the opacity at \SI{850}{\micro\metre} and the extinction coefficient in the $K_S$ passband (\SI{{\sim}2.2}{\micro\metre}). In light of our findings with respect to cloud substructure and the consequential introduced bias in the slope, we conclude, that at the moment it is difficult to make decisive physical interpretations, given a particular value of $\gamma$. More studies in regions, where extinction maps convincingly sample the entire cloud structure are necessary to arrive at a better understanding of the slope $\gamma$. Similar to what we found with the new VISTA data for \oriona, future observations will provide better insight into this problem. In particular, the VISTA Star Formation Atlas \footnote{VISIONS - The VISTA Star Formation Atlas is an ongoing ESO public survey and is based on this paper series. More information is available at \href{http://visions.univie.ac.at}{http://visions.univie.ac.at}.} will performs high-sensitivity, wide-field $JHK_S$ observations of major nearby star-forming complexes. The NIR photometry will enable the construction of high resolution extinction maps, with well-sampled pixels and therefore will provide further insight into this problem.

\section{Summary}
\label{sec:summary}

In this second publication in the context of the Vienna Survey in Orion, we have investigated infrared extinction properties in the \oriona molecular cloud complex. The main results of this paper are as follows.

\begin{enumerate}

    \item We combined several large-scale photometric databases, covering a wavelength range from just below \SI{1}{\um} to about \SI{25}{\um}. Our master catalogue includes NIR data from Pan-STARRS, as well as HAWK-I and the VISTA observations presented in \citetalias{meingast16}. The MIR is covered by \textit{Spitzer} IRAC and MIPS photometry. A detailed description of our final catalogue is given in Sect. \ref{sec:data}.

    \item Our results critically depend on the reliability of linear fitting procedures. For this reason, we have developed a two-dimensional Bayesian framework which is used together with a Markov chain Monte Carlo algorithm. Specifically, we have derived a model likelihood which takes measurement errors in both dimensions into account. This framework can be adapted to include more complex models and therefore can be used in a multitude of additional aspects far beyond the general scope of this publication. The final model likelihood is given in Equ. \ref{equ:likelihood}, for which a detailed derivation can be found in Sect. \ref{sec:methods:fitting}.

    \item We derived the infrared extinction law in \oriona by fitting colour excess ratios in a variety of colour-colour diagrams. Here, we first measured the average extinction law (Sect. \ref{sec:results:average}), and continued to investigate spatial variations across the molecular cloud complex (Sect. \ref{sec:results:spatial}). Fitting results and the total extinction ratios for all included passbands are available in Tab.~\ref{tab:law}. The results for the fitting procedures are visualized for the $H$ passband in Figs. \ref{img:total_H} and \ref{img:split_H}. For the average extinction law we find excellent agreement compared to results in the literature.

    \item For the passbands up to about \SI{6}{\um} we find statistically significant spatial variations in the colour excess ratios of the order of 3\%. We investigated the origin of these variations by splitting the survey into distinct sub-regions (Fig.~\ref{img:extinction_law_regions}). Here, we find a clear signature, that those regions, which seem to be affected by radiative feedback from the massive stars in the ONC, show a different extinction law when compared to other star-forming sites in the cloud (Fig.~\ref{img:beta_subregions}).

    \item To support and interpret our results on the spatial variations from the linear fitting procedure, we compared our findings to various model predictions (Fig.~\ref{img:extinction_law}). Here, we found that the general trends can be reproduced by the models, reinforcing our claim with respect to a variable extinction law. However, a more detailed analysis revealed that these models are not able to fully explain all details of our findings across the entire infrared spectral range.

    \item The \oriona MIR extinction law is characteristically flat, similar to many other recently investigated sightlines (Fig.~\ref{img:extinction_law}). This result contradicts previous investigations in the \oriona region. Furthermore, flat MIR extinction is oftentimes associated with extraordinary large dust grains (\SI{{\sim}10}{\um}). However, a comparison with recently published dust grain models shows that this must not necessarily be the case (Sect. \ref{sec:results:law:discussion}).

    \item Based on the derived average extinction law, we constructed a new extinction map of \oriona with a resolution of \SI{1}{\arcmin} (Fig.~\ref{img:extinction_map}). We compared this map to \textit{Herschel} dust optical depth measurements (Sect. \ref{sec:results:herschel}) and derived a new calibration to convert these data to extinction (Fig.~\ref{img:herschel_calibration}). However, we also found alarming evidence, that the conversion of dust optical depth, as derived from \textit{Herschel} data, to column densities is likely biased by cloud substructure which is not sampled by background sources (Fig.~\ref{img:substructure_maglim}). Another source of systematic uncertainty in this correlation may be the conversion of the dust emission measurements to dust optical depth itself.

\end{enumerate}
Finally, we want to highlight that \oriona presents a unique opportunity to study the extinction law in an unbiased way. Many well-known results in the literature are based on measurements in the galactic plane and specifically towards the galactic centre. In these regions, however, it is difficult to interpret the extinction law and its variations, since many physically separate layers of interstellar material can overlap along the line of sight. Therefore, and in light of our findings on the variations in the extinction law, we argue that it is extremely difficult to find such small deviations (3\%) in these regions. Moreover, since \oriona hosts the nearest massive star-forming region, it is the ideal location for detailed investigations into dust grain populations and their generic properties under the influence of radiative feedback.

\begin{acknowledgements}
We want to thank the anonymous referee for the very detailed and useful comments that helped to improve the quality of this publication.
Based on observations collected at the European Organisation for Astronomical Research in the Southern Hemisphere.
under ESO programmes 082.C-0032(A) and 090.C-0797(A).
This research made use of Astropy, a community-developed core Python package for Astronomy \citep{astropy}.
This research has made use of "Aladin sky atlas" developed at CDS, Strasbourg Observatory, France \citep{bonnarel00}.
This research has made use of the SVO Filter Profile Service (\href{http://svo2.cab.inta-csic.es/theory/fps/}{http://svo2.cab.inta-csic.es/theory/fps/}) supported from the Spanish MINECO through grant AyA2014-55216.
We also acknowledge the various Python packages that were used in the data analysis of this work, including NumPy \citep{numpy}, SciPy \citep{scipy}, scikit-learn \citep{scikit-learn}, and Matplotlib \citep{matplotlib}.
This research has made use of the SIMBAD database operated at CDS, Strasbourg, France \citep{simbad}.
\end{acknowledgements}

\bibliography{bibliography}

\end{document}